\documentclass{statsoc}
\usepackage{amsmath,amssymb,multirow,multicol,natbib}
\usepackage{graphicx}
\usepackage{rasmus2}

\def \bu{{\mathbf u}}
\def \bv{{\mathbf v}}
\def \br{{\mathbf r}}
\def \bz{{\mathbf z}}
\def \bx{{\mathbf x}}
\def \be{{\mathbf e}}
\def \bI{{\mathbf I}}
\def \bJ{{\mathbf J}}
\def \bS{{\mathbf S}}
\def \bff{{\mathbf f}}
\def \bT{{\mathbf T}}
\def \bY{{\mathbf Y}}
\def \bV{{\mathbf V}}
\def \bD{{\mathbf D}}
\def \bG{{\mathbf G}}
\def \bL{{\mathbf L}}
\def \bA{{\mathbf A}}
\def \bM{{\mathbf M}}
\def \bbt{{\boldsymbol{\beta}}}
\def \bsig{{\boldsymbol{\Sigma}}}
\def \bphi{{\boldsymbol{\phi}}}
\def \blambda{{\boldsymbol{\lambda}}}
\def \bpsi{{\boldsymbol{\psi}}}
\def \btheta{{\boldsymbol{\theta}}}
\def \bmu{{\boldsymbol{\mu}}}
\def \beq{\begin{equation}}
\def \eeq{\end{equation}}

\title[Quasi-likelihood for Spatial Point Processes]{Quasi-likelihood for Spatial Point Processes}
\author[Guan {\it et al.}]{Yongtao Guan}
\address{Miami, USA}
\author{Abdollah Jalilian}
\address{Kermanshah, Iran}
\author{Rasmus Waagepetersen}
\coaddress{Department of Mathematical Sciences, Aalborg University,
	   Fredrik Bajersvej 7G, DK-9220 Aalborg, Denmark}
\address{Aalborg, Denmark}

\email{rw@math.aau.dk}
\date{}

\begin{document}

\maketitle



\begin{abstract}

Fitting regression models for intensity functions of spatial point processes is of great interest in ecological and epidemiological studies of association between spatially referenced events and geographical or environmental covariates. When Cox or cluster process models are used to accommodate clustering not accounted for by the available covariates, likelihood based inference becomes computationally cumbersome due to the complicated nature of the likelihood function and the associated score function. It is therefore of interest to consider alternative more easily computable estimating functions. We
derive the optimal estimating function in a class of first-order estimating functions.
The optimal estimating function depends on the solution of a certain Fredholm integral
equation which in practice is solved numerically. The approximate solution
is equivalent to a quasi-likelihood for binary spatial data and we
therefore use the term quasi-likelihood for our optimal estimating
function approach. We demonstrate in a simulation
study and a data example that our quasi-likelihood method for spatial
point processes is both statistically and computationally efficient.
\end{abstract}

\section{INTRODUCTION}

In many applications of spatial point processes it is of interest to
fit a regression model for the intensity function. In case of a
Poisson point process, maximum likelihood estimation of regression
parameters is rather straightforward with a user-friendly
implementation available in the \texttt{R} package
\texttt{spatstat}. However, if Cox or cluster point process models are
used to accommodate clustering not explained by a Poisson process,
then maximum likelihood estimation is in general difficult from a
computational point of view (see e.g.\ M{\o}ller and Waagepetersen, 2004)\nocite{moeller:waagepetersen:04}. Alternatively, one may follow composite likelihood arguments \cite[e.g.][]{moeller:waagepetersen:07}
to obtain an estimating function that is equivalent to the score of the
Poisson likelihood function. This
provides a computationally tractable estimating function and theoretical properties of the
resulting estimator are well understood, see e.g.\ \cite{schoenberg:05},
\cite{waagepetersen:07} and \cite{guan:loh:07}.

A drawback of the Poisson score function approach is the loss of efficiency
since possible dependence between points is ignored. In the context of
intensity estimation, it appears that only \cite{mrkvicka:molchanov:05}
and \cite{guan:shen:10} have tried to incorporate second-order properties
in the
estimation so as to improve efficiency.  \cite{mrkvicka:molchanov:05} show
that their proposed estimator is optimal among a class of linear, unbiased
intensity estimators, where the word `optimal' refers to minimum variance.
However, their approach is confined to a very restrictive type of intensity function known up to a one-dimensional scaling factor. In contrast, \cite{guan:shen:10} propose a weighted estimating
equation approach that is applicable to intensity functions in more general
forms. A similar optimality result can on the other hand not be established for their
approach.

In this paper we derive an optimal estimating function that not only
takes into account possible spatial correlation but also is applicable
for point processes with a general regression model for the intensity
function. In the spirit of generalized linear models the intensity is
given by a differentiable function of a linear predictor depending on
spatial covariates. The optimal estimating function depends on the
solution of a certain Fredholm integral equation and reduces to the
likelihood score in case of a Poisson process. We show in Section~3.2
that the optimality result in \cite{mrkvicka:molchanov:05} is a
special case of our more general result, and that the estimation
method in \cite{guan:shen:10} is only a crude approximation of our new
approach. Apart from being computationally efficient, our estimating
function only requires specification of the intensity function and the
so-called pair correlation function, which is another advantage compared with maximum likelihood estimation.

For many types of correlated data other than spatial point patterns,
estimating functions have been widely used for model fitting
when maximum likelihood estimation is computationally challenging.
Examples of such data include longitudinal data \cite[][]{liang:zeger:86}, time
series data \cite[][]{zeger:88}, clustered failure time data \cite[][]{gray:03}
and spatial binary or count data \cite[][]{gotway:stroup:97,lin:clayton:05}.
For most of these methods, the inverse of a covariance matrix is used in
their formulations as a way to account for the correlation in data, and
optimality can be established when the so-called quasi-score estimating
functions are used \cite[][]{heyde:97}. For point processes there is not a direct analogue of a spatial covariance matrix, but it turns out that a numerical implementation of our method is closely related to the quasi-likelihood
for spatial data considered in \cite{gotway:stroup:97} and \cite{lin:clayton:05}.  Our work hence not only lays the theoretical foundation for optimal intensity estimation, but also fills in a critical gap between existing literature on spatial
point processes and the well-established quasi-likelihood estimation method. We therefore adopt the term quasi-likelihood for our approach.

Following some background material on point processes and estimating functions, we derive our optimal estimating function and discuss the practical implementation of it based on a numerical solution of the Fredholm integral equation. Asymptotic properties of the resulting parameter estimator is then considered and the superior performance of the quasi-likelihood method compared with existing ones is demonstrated through a simulation study. We finally illustrate the practical use of the quasi-likelihood in a data example of three tropical tree species.

\section{BACKGROUND}\label{sec:background} 

In this section we provide background on first- and second-order moments of spatial point processes, composite likelihood estimation and estimating functions. Throughout the presentation, we use $\EE$, $\Var$ and $\Cov$ to denote expectation, variance and covariance, respectively.

\subsection{Intensity and Pair Correlation Function}\label{sec:intensityandpcf}

Let $X$ be a point process on $\R^2$ and let $N(B)$ denote the number of points
in $X \cap B$ for any bounded (Borel) set $B\subseteq \R^2$. We assume that $X$ has an intensity function $\lambda(\cdot)$ and a pair correlation function $g(\cdot,\cdot)$, whereby the first- and second-order moments
of the counts $N(B)$ are given by
\beq \label{eq:meancount}
  \EE N(B) = \int_B \lambda(\bu) \dd \bu
\eeq
and
\begin{equation}\label{eq:countcovariance}
  \Cov[ N(A), N(B)] = \int_{A \cap B} \lambda(\bu)\dd\bu
   + \int_A \int_B \lambda(\bu)\lambda(\bv)[ g(\bu,\bv)-1] \dd\bu \dd\bv 
\end{equation}
for bounded sets $A,B \subseteq \R^2$ \citep{moeller:waagepetersen:04}.

For convenience of exposition we assume that $g(\bu, \bv)$ only depends on the
difference $\bu - \bv$ since this is the common assumption in practice. In the
following we thus let $g(\br)$ denote the pair correlation function for two points
$\bu$ and $\bv$ with $\bu - \bv=\br$. However, our proposed optimal estimating
function is applicable also in the case of a non-translation invariant pair
correlation function.

\subsection{Composite Likelihood}\label{sec:cl}

Assume that the intensity function is given in terms of a parametric model
$\lambda(\bu)=\lambda(\bu;\bbt)$, where $\bbt =(\bt_1,\ldots,\bt_p) \in \R^p$
is a vector of regression parameters. Popular choices of the parametric model include linear and
log linear models, $\lambda(\bu;\bbt) = \bz(\bu)\bbt^\T$ and
$\log\lambda(\bu;\bbt)=\bz(\bu)\bbt^\T$, where $\bz(\bu) = (z_1(\bu),\ldots,z_p(\bu))$
is a  covariate vector for each $\bu \in \R^2$. A first-order log composite likelihood
function \cite[][]{schoenberg:05,waagepetersen:07} for estimation of $\bbt$
is given by
\begin{equation}\label{eq:cl}
  \sum_{\bu \in X \cap W} \log \lambda(\bu; \bbt) - \int_W \lambda(\bu;\bbt) \dd \bu,
\end{equation}
where $W \subset \R^2$ is the observation window. This can be viewed as a limit
of log composite likelihood functions for binary variables $Y_i=1[N(B_i)>0]$,
$i=1,\ldots,m$, where the cells $B_i$ form a disjoint partitioning of $W$ and
$1[\cdot]$ is an indicator function \cite[e.g.][]{moeller:waagepetersen:07}.
The limit is obtained when the number of cells tends to infinity and the areas of
the cells tend to zero. In case of a Poisson process, the composite likelihood coincides with the likelihood function.

The composite likelihood is computationally simple and enjoys considerable
popularity in particular in studies of tropical rain forest ecology where spatial
point process models are fitted to huge spatial point pattern data sets of rain forest
tree locations \cite[see e.g.][]{shen:et:al:09,lin:et:al:11}. However, it is not
statistically efficient for non-Poisson data since possible correlations between counts
of points are ignored.


\subsection{Primer on Estimating Functions}\label{sec:eeprimer}

Referring to the previous Section~\ref{sec:cl}, the composite likelihood estimator of
$\bbt$ is obtained by maximizing the log composite likelihood \eqref{eq:cl}. Assuming that $\lambda$ is differentiable with respect to $\bbt$ with gradient $\blambda'(\bu;\bbt) = \dd \lambda(\bu;\bbt)/\dd \bbt$, this is equivalent to solving the following equation:
\begin{equation}\label{eq:ee} \be (\bbt)= \mathbf{0},
\end{equation}
where
 \begin{equation}\label{eq:poislikescore}
   \be (\bbt)= \sum_{\bu \in X \cap W} \frac{\blambda'(\bu;\bbt)}{\lambda(\bu;\bbt)}
    - \int_W \blambda'(\bu;\bbt) \dd \bu
 \end{equation}
is the gradient of \eqref{eq:cl} with respect to $\bbt$. Equations in the form of \eqref{eq:ee} are typically referred to as estimating equations and functions like $\be (\bbt)$ are called estimating functions \citep{heyde:97}.
Note that many other statistical estimation procedures, such as maximum likelihood estimation, moment based estimation and minimum contrast estimation, can all be written in terms of estimating functions. 

We defer rigorous asymptotic details to Section~\ref{sec:asymptotics} and here just provide an informal overview of properties of an estimator $\hat \bbt$ based on an estimating function $\be(\bbt)$. By a first-order Taylor series expansion at $\hat \bbt$,
\[ \be(\bbt) \approx \be (\hat \bbt)+[ \hat \bbt - \bbt]\bS =  ( \hat \bbt - \bbt)\bS, \]
where $\bS = -  \EE {\dd \be(\bbt)}/{\dd \bbt^\T}$
is the so-called sensitivity matrix \cite[e.g.\ page 62 in][]{song:07} and the equality is due to $\be (\hat \bbt)=\mathbf{0}$ as required by \eqref{eq:ee}. It then follows immediately that $ \hat \bbt \approx \bbt +  \be(\bbt) \bS^{-1} $.
Thus, with $\bbt$ equal to the true parameter value, $\hat \bbt$ is approximately unbiased if $\EE \be(\bbt)=0$, i.e.\ $\be(\bbt)$ is an unbiased estimating function. Moreover, $\Var \hat \bbt \approx \bS^{-1} \bsig \bS^{-1} $
where $\bsig=\Var \be(\bbt)$ and $\bS^{-1} \bsig \bS^{-1}$ is  the asymptotic covariance matrix when the size of the data set goes to infinity in a suitable manner (Section~\ref{sec:asymptotics}). The inverse of $\bS^{-1} \bsig \bS^{-1}$, i.e. $\bS \bsig^{-1} \bS$, is called the Godambe information \cite[e.g.\ Definition 3.7 in][]{song:07}.

Suppose that two competing estimating functions $\be_1(\bbt)$ and $\be_2(\bbt)$ with respective Godambe informations $\bI_1$ and $\bI_2$ are used to obtain the estimators $\hat \bbt_1$ and $\hat \bbt_2$. Then $\be_1(\bbt)$ is said to be superior to $\be_2(\bbt)$ if $\bI_1-\bI_2$ is positive definite, since this essentially means that $\hat \bbt_1$ has a smaller asymptotic variance than $\hat \bbt_2$. If $\bI_1-\bI_2$ is positive definite for all possible $\be_2(\bbt)$, then we say that $\be_1(\bbt)$ has the maximal Godambe information and is an optimal estimating function. The resulting estimator $\hat\bbt_1$ is then the asymptotically most efficient.

\section{AN OPTIMAL FIRST-ORDER ESTIMATING EQUATION}\label{sec:gee}

The estimating function given in \eqref{eq:poislikescore} can be rewritten as
\beq\label{eq:geef}
  \be_\bff(\bbt) = \sum_{\bu \in X \cap W} \bff(\bu) - \int_W\bff(\bu)\lambda(\bu;\bbt)\dd\bu,
\eeq
where $\bff(\bu)=\blambda'(\bu;\bbt)/\lambda(\bu;\bbt)$. In general, $\bff(\bu)$ can be any $1 \times p$ real vector valued function, where $p$ is the dimension of $\bbt$. We call \eqref{eq:geef}  a first-order estimating function. Our aim is to find a function $\bphi$ so that $\be_{\bphi}$ is optimal within the class of first-order estimating functions; in other words, the resulting estimator of $\bbt$ associated with $\be_{\bphi}$ is asymptotically most efficient. 

Let $\bsig_\bff= \Var \be_\bff(\bbt)$, $\bJ_\bff= - \dd \be_\bff(\bbt)/ \dd \bbt^\T$
and $\bS_\bff=\EE \bJ_\bff$. Note that $\bsig_\bff$, $\bJ_\bff$ and $\bS_\bff$ all depend on $\bbt$ but we suppress this dependence in this section for ease of presentation. Recalling the definition of optimality in Section~2.3, for $\be_{\bphi}$ to be optimal we must have that
\beq \label{eq:godambediff}
  \bS_\bphi \bsig_\bphi^{-1} \bS_\bphi - \bS_\bff \bsig_\bff^{-1} \bS_\bff
\eeq
is non-negative definite for all $\bff:W\to\R^p$. A sufficient condition for
this is
\begin{equation}\label{eq:sufficient} \bsig_{\bphi \bff} = \bS_\bff \
\end{equation}
for all $\bff$ where $\bsig_{\bff\bphi} = \Cov[\be_\bff(\bbt), \be_\bphi(\bbt)]$. This type of condition is provided in Theorem~2.1 in \cite{heyde:97} for  discrete or continuous vector-valued data. In Appendix~\ref{app:optimality}, we give a short self-contained proof of the sufficiency of \eqref{eq:sufficient} in our setting.

By the Campbell formulae \cite[e.g.][Chapter~4]{moeller:waagepetersen:04},
\begin{align*}
\bsig_{\bphi \bff} &= \! \! \int_W \! \! \! \bff^\T(\bu) \bphi(\bu) \lambda(\bu;\bbt) \dd \bu
  + \! \int_{W^2} \! \! \! \!\!
  \bff^\T(\bu) \bphi(\bv) \lambda(\bu;\bbt) \lambda(\bv;\bbt)[g(\bu - \bv) - 1] \dd \bu \dd \bv, \nonumber \\ 
  \bS_\bff &= \int_W \bff^\T(\bu) \blambda'(\bu;\bbt) \dd\bu.
\end{align*}
Hence, \eqref{eq:sufficient} is equivalent to
\[
  \int_W \bff^\T(\bu) \big\{ \blambda'(\bu;\bbt) - \bphi(\bu) \lambda(\bu;\bbt) -
    \lambda(\bu;\bbt) \int_W \phi(\bv) \lambda(\bv;\bbt)[g(\bu - \bv) - 1]  \dd \bv \big\} \dd \bu = {\bf 0}
\]
for all $\bff:W\to\R^p$, which is true if
\begin{equation}\label{eq:justification}
\blambda'(\bu;\bbt) - \bphi(\bu) \lambda(\bu;\bbt) -
    \lambda(\bu;\bbt) \int_W \phi(\bv) \lambda(\bv;\bbt)[g(\bu - \bv) - 1]  \dd \bv=0
\end{equation}
for all $\bu\in W$. Assuming $\lambda>0$, \eqref{eq:justification} implies that $\bphi$ is a solution to the Fredholm integral
equation \cite[e.g.][Chapter~3]{hackbusch:95}
\beq \label{eq:fredholm}
  \bphi = \frac{\blambda'}{\lambda} - \bT \bphi,
\eeq
where $\bT$ is the operator given by
\beq \label{eq:T}
 (\bT\bff)(\bu)= \int_W t(\bu,\bv)\bff(\bv)  \dd \bv \ \text{ with } \
  t(\bu,\bv) = \lambda(\bv;\bbt)[g(\bu - \bv) - 1].
\eeq

Assume that $g$ is continuous so that $\bT$ is compact in the space of
continuous functions on $W$ \cite[Theorem 3.2.5]{hackbusch:95} and moreover that $-1$ is not an eigenvalue (we return to this condition in the next section). It then follows by Theorem~3.2.1 in \cite{hackbusch:95} that
\eqref{eq:fredholm} has a unique solution
\[
   \bphi = (\bI + \bT)^{-1}  \frac{\blambda'}{\lambda},
\]
where $\bI$ is the identity operator (or, depending on context, the identity matrix) and $(\bI + \bT)^{-1}$ is the bounded linear
inverse of $\bI + \bT$.
We define
\begin{align}\label{eq:limitgee}
   \be(\bbt) & = \be_\bphi(\bbt) = \sum_{\bu \in X\cap W} \bphi(\bu)
    - \int_{W} \bphi(\bu) \lambda(\bu;\bbt) \dd\bu, \\
\bsig & = \Var \be(\bbt), \, \bJ = - \dd \be(\bbt) / \dd \bbt^\T, \, \bS = \EE \bJ \nonumber
\end{align}
where by the above derivations,
\beq \label{eq:SandSigma}
 \bS  = \bsig = \int_W \bphi^\T(\bu) \blambda'(\bu;\bbt) \dd\bu.
\eeq
In the Poisson process case where $g(\cdot) = 1$, \eqref{eq:limitgee} reduces to
the Poisson likelihood score \eqref{eq:poislikescore}.

We develop a more explicit expression for $\bphi$ by using Neumann series expansion in Appendix~\ref{sec:neumanncondition}. The Neumann series expansion is also useful for checking the conditions for our asymptotic results; see Appendix~\ref{sec:conditions}. However, it is not essential for our approach so we omit the detailed discussion here.
\subsection{Condition for non-negative eigenvalues of $\bT$}\label{sec:nonnegeigen}

In general it is difficult to assess the eigenvalues of $\bT$ given by \eqref{eq:T}. However, suppose that $g - 1$ is non-negative definite so that $\bT^s$ is a positive operator (i.e., $\int_W \bff^{\T}(\bu)(\bT^s \bff)(\bu) \dd\bu \geq 0$) where $\bT^s$ is given by the symmetric kernel
\[ t^s(\bu, \bv) = \lambda(\bu;\bbt)^{1/2}  \lambda(\bv;\bbt)^{1/2} \big[ g(\bu - \bv) -1\big].
\] Then all eigenvalues of $\bT^s$ are non-negative \cite[][Corollary 1, p. 320]{lax:02}. In particular, $-1$ is not an eigenvalue. The same holds for $\bT$ since it is easy to see that the eigenvalues of $\bT$ coincide with those of $\bT^s$. 

The assumption of a non-negative definite $g(\cdot)-1$ is valid for the wide class of Cox point processes which in turn includes the class of Poisson cluster processes. For a Cox process driven by a random intensity function $\Ld$,
$g(u,v)=1+\Cov[\Ld(\bu),\Ld(\bv)]/[\lambda(\bu)\lambda(\bv)]$ so that
$g(\cdot)-1$ is non-negative definite.

\subsection{Relation to Existing Methods}\label{sec:weightedlikelihood}

Suppose  we approximate the operator $\bT$ by
\begin{equation}\label{eq:wclapprox}
   (\bT\bff)(\bu) = \int_W \bff(\bv) \lambda(\bv;\bbt) [ g(\bu - \bv)-1)] \dd\bv
   \approx \lambda(\bu;\bbt) \bff({\bf u}) \int_W[ g(\bu - \bv)-1] \dd \bv.
\end{equation}
This is justified if $\bff(\bv) \lambda(\bv;\bbt)$ is close to $\bff(\bu)\lambda(\bu;\bbt)$
for the $\bv$ where $g(\bu - \bv) - 1$ differs substantially from zero. Then
the Fredholm integral equation \eqref{eq:fredholm} can be approximated by
\[
   \bphi = \frac{\blambda'}{\lambda}  - \lambda A \bphi,
\]
where
\[
  A(\bu) = \int_{W} \big[g(\bu - \bv) - 1\big] \dd\bv.
\]
We hence obtain an approximate solution $\bphi = w \blambda'/\lambda$ with
$w(\bu) = [1 + \lambda(\bu;\bbt) A(\bu)]^{-1}$.
Using this approximation in \eqref{eq:limitgee} we obtain the estimating function
\[
   \sum_{\bu \in X\cap W} w(\bu)\frac{\blambda'(\bu;\bbt)}{\lambda(\bu;\bbt)}
   - \int_{W} w(\bu) \blambda'(\bu;\bbt) \dd\bu,  \]
which is precisely the weighted Poisson score suggested in \cite{guan:shen:10}.

\cite{mrkvicka:molchanov:05} derived optimal intensity estimators in the
situation of $\lambda(\bu;\rho) = \rho \gamma(\bu)$ for some known function
$\gamma(\bu)$ and unknown parameter $\rho>0$. Since $\rho$ is the only
unknown parameter, a direct application of \eqref{eq:fredholm} yields
\[
   \rho\bphi(\bu) + \rho^2 \int_W \bphi(\bv)\gamma(\bv) \big[g(\bu - \bv) - 1\big]\dd\bv = 1,
\]
which is essentially Corollary 3.1 of \cite{mrkvicka:molchanov:05}. It is uncommon for an intensity function to be known up to a one-dimensional scaling factor. In contrast, our proposed modeling framework for the intensity function closely mimics that used in classical regression analysis and is more general. As a result, our method of derivation is completely different from that in \cite{mrkvicka:molchanov:05}.

\section{IMPLEMENTATION}\label{sec:implementation}

In this section we discuss practical issues concerning the implementation
of our proposed optimal estimating function. In particular we show in Section~\ref{sec:ql} that a particular numerical approximation of our optimal estimating function is equivalent to a quasi-likelihood for binary spatial data for which an iterative generalized least squares solution can be implemented. An \texttt{R} implementation will appear in future releases of \texttt{spatstat}.

\subsection{Numerical Approximation}\label{sec:numerical}

To estimate $\bphi$, consider the numerical approximation
\begin{equation}\label{eq:quadrature}
  (\bT\bphi)(\bu) = \int_W t(\bu,\bv) \bphi(\bv) \dd \bv \approx \sum_{i=1}^m t(\bu,\bu_i) \bphi(\bu_i) w_i,
\end{equation}
where $\bu_i, i=1,\ldots,m$, are quadrature points with associated
weights $w_i$. Inserting this approximation in \eqref{eq:fredholm}
with $\bu=\bu_l$
we obtain estimates $\hat\bphi(\bu_l)$ of $\bphi(\bu_l)$,
$l=1,\ldots,m$, by solving the system of linear equations,
$$
  \bphi(\bu_l) + \sum_{i=1}^m t(\bu_l, \bu_i) \bphi(\bu_i)w_i
   = {\blambda'(\bu_l;\bbt)\over\lambda(\bu_l;\bbt)},\;\; l=1,\ldots,m.
$$
Then $(\bT\bphi)(\bu) \approx \sum_{i=1}^m t(\bu,\bu_i) \hat
\bphi(\bu_i) w_i$ and plugging this further approximation into \eqref{eq:fredholm},
the Nystr{\" o}m approximate solution of \eqref{eq:fredholm} directly becomes
\beq \label{eq:dfred}
   \hat\bphi(\bu) = {\blambda'(\bu;\bbt)\over\lambda(\bu;\bbt)}
   - \sum_{i=1}^m t(\bu,\bu_i) \hat\bphi(\bu_i) w_i.
\eeq
In \eqref{eq:limitgee} we replace $\bphi $ by $\hat\bphi$ and we approximate the integral term applying again the quadrature rule used to obtain $\hat\bphi$. This leads to
\begin{equation}\label{eq:nysgee}
  \hat \be(\bbt) = \sum_{\bu\in X\cap W} \hat\bphi(\bu)
   - \sum_{i=1}^m \hat\bphi(\bu_i)\lambda(\bu_i;\bbt) w_i.
\end{equation}

To estimate $\bbt$, we solve $\hat \be(\bbt)=0$ iteratively using Fisher scoring.
Suppose that the current estimate is $\bbt^{(l)}$. Then $\bbt^{(l+1)}$ is obtained
by the Fisher scoring update
\beq \label{eq:fisherscoring}
  \bbt^{(l+1)} = \bbt^{(l)} + \hat \be(\bbt^{(l)}) \hat \bS^{-1},
\eeq
where
\begin{equation}\label{eq:Shat}
  \hat \bS = \sum_{i=1}^m \hat\bphi(\bu_i)^\T \blambda'(\bu_i;\bbt^{(l)}) w_i
\end{equation}
is the numerical approximation of the sensitivity matrix
$\bS = \int_{W} \bphi^\T(\bu)\blambda'(\bu;\bbt^{(l)}) \dd \bu$.


Provided the quadrature scheme is convergent, it follows by Lemma~4.7.4, Lemma~4.7.6
and Theorem~4.7.7 in~\cite{hackbusch:95} that $\|\bphi - \hat \bphi\|_{\infty}$
converges to zero as $m\to\infty$. This justifies the use of the Nystr{\"o}m
method to obtain an approximate solution of the Fredholm integral equation.

\subsection{Implementation as quasi-likelihood}\label{sec:ql}

Suppose that we are using  
simple Riemann quadrature in \eqref{eq:quadrature}. Then the $w_i$'s correspond
to areas of some sets $B_i$ that partition $W$ and for each $i$, $\bu_i \in B_i$.
Let $Y_i$ denote the number of points from $X$ falling in $B_i$ and define $\mu_i = \lambda(\bu_i;\bbt) w_i$.
If the $B_i$'s are sufficiently small so that the $Y_i$'s are binary then \eqref{eq:nysgee} is approximately equal to
\beq \label{eq:nysgee2}
   \sum_{i=1}^m \hat\bphi(\bu_i) (Y_i - \mu_i).
\eeq
Further,  by \eqref{eq:meancount} and \eqref{eq:countcovariance}, $\EE Y_i \approx \mu_i$ and
\begin{align*}
   \Cov( Y_i, Y_j ) &= 1(i = j)\int_{B_i} \lambda(\bu;\bbt) \dd\bu
    + \int_{B_i \times B_j} \lambda(\bu;\bbt)\lambda(\bv;\bbt) \big[ g(\bu - \bv) - 1 \big] \dd \bu \dd \bv  \\
   &\approx  V_{ij} = \mu_i 1(i=j) + \mu_i \mu_j \big[ g(\bu_i, \bu_j) - 1 \big].
\end{align*}
Define $\bY=(Y_i)_i$, $\bmu=(\mu_i)_i$ and $\bV = [V_{ij}]_{ij}$.
Then $\EE \bY \approx \bmu$ and $ \Cov \bY \approx \bV$.
Moreover, from \eqref{eq:dfred}, $[\hat \bphi(\bu_i)]_i = \bV^{-1}\bD$ where
$\bD = \dd \bmu^\T / \dd \bbt$ is the $m \times p$ matrix of partial derivatives
$\dd \mu_i/\dd \beta_j$. Hence, \eqref{eq:nysgee2} becomes
\begin{equation}\label{eq:discrete}
   (\bY - \bmu) \bV^{-1} \bD,
\end{equation}
which is formally a quasi-likelihood score for spatial data $\bY$ with mean $\bmu$
and covariance matrix $\bV$ \cite[][]{gotway:stroup:97}.

Similarly,  $\hat \bS$ in \eqref{eq:Shat} becomes $\bD^\T \bV^{-1} \bD$ and substituting $\hat \be$ in  \eqref{eq:fisherscoring}  by \eqref{eq:discrete}, we obtain the iterative generalized least
squares equation
\begin{equation}\label{eq:igls}
   (\bbt^{(l+1)} - \bbt^{(l)}) \bD(\bbt^{(l)})^\T \bV(\bbt^{(l)})^{-1} \bD(\bbt^{(l)})
   = [\bY - \bmu(\bbt^{(l)})] \bV(\bbt^{(l)})^{-1}  \bD(\bbt^{(l)}),
\end{equation}
where we have used the notation $\bD(\bbt)$, $\bV(\bbt)$ and $\bmu(\bbt)$ to
emphasize the dependence of $\bD$, $\bV$, and $\bmu$  on $\bbt$.

\subsection{Preliminary Estimation of Intensity and Pair Correlation}\label{sec:gandtilderho}

Using the notation from Section~\ref{sec:ql}, $\bV = \bV_\bmu^{1/2} (\bI + \bG) \bV_\bmu^{1/2}$
where $\bV_\bmu = \text{Diag}(\mu_i)$ and
\[
   G_{ij} = \sqrt{\mu_i \mu_j}\big[ g(\bu_i,\bu_j) - 1 \big]
\]
so that $\bG=[G_{ij}]_{ij}$ is the matrix analogue of the symmetric operator $\bT^s$ from
Section~\ref{sec:nonnegeigen}. In general $g$ is unknown and must be replaced by
an estimate. Moreover it is advantageous if $\bG$ is fixed in order to avoid the computational burden of repeated matrix
inversion in the generalized least squares iterations \eqref{eq:igls}.

To estimate $g$ we assume that $g(\br) = g(\br;\bpsi)$ where $g(\cdot;\bpsi)$
is a translation invariant parametric pair correlation function model. We
replace $\bpsi$ and $\bbt$ inside $\bG$ by preliminary estimates $\tilde \bbt$ and $\tilde \bpsi$ which are fixed during the  iterations \eqref{eq:igls}. The
estimates $\tilde \bbt$ and $\tilde \bpsi$ can be obtained using the two-step
approach in~\cite{waagepetersen:guan:09} where $\tilde \bbt$ is obtained from
the composite likelihood function and $\tilde \bpsi$ is a minimum contrast
estimate based on the $K$-function. If translation invariance can not be assumed,
$\bpsi$ may instead be estimated by using a second-order composite likelihood
as in \cite{jalilian:guan:waagepetersen:10}.

\subsection{Tapering}\label{sec:tapering}

The matrix $\bV$ can be of very high dimension. However, many entries in
$\bV$ are very close to zero and we can therefore approximate $\bV$ by a
sparse matrix $\bV_\text{taper}$ obtained by tapering \cite[e.g.][]{furrer:genton:nychka:06}.
More precisely, we replace $\bG$ in $\bV$ by a matrix $\bG_{\text{taper}}$
obtained by assigning zero to entries $G_{ij}$ below a suitable threshold.
We then compute a sparse matrix Cholesky decomposition, $\bI + \bG_\text{taper} = \bL\bL^\T$.
Then $(\bY - \bmu)\bV^{-1/2}_\bmu (\bI + \bG_\text{taper})^{-1}$ can be
easily computed by solving the equation $\mathbf{x} \bL\bL^\T = (\bY - \bmu) \bV_\bmu^{-1/2}$
in terms of $\mathbf{x}$ using forward and back substitution
for the sparse Cholesky factors $\bL$ and $\bL^\T$, respectively.

In practice, it is often assumed that $g({\bf r}) = g_0(\|{\bf r}\|)$ for some function $g_0$. If $g_0$ is a decreasing function of $\|{\bf r}\|$ then we may define the entries in $\bG_\text{taper}$ as
$G_{ij}1[\|\bu_i - \bu_j\| \leq d_\text{taper}]$, where $d_\text{taper}$ solves
$[g_0(d)-1]/[g_0(0)-1] = \epsilon$ for some small $\epsilon$. That is, we replace entries $G_{ij}$ by zero if $g_0(\|\bu_i - \bu_j\|)-1$ is below some small percentage of
the maximal value $g_0(0)-1$.

When $\bV$ in~\eqref{eq:igls} is replaced by $\bV_\text{taper}$ we obtain the following
estimate of the covariance matrix of $\hat \bbt$:
\beq \label{eq:tapercov}
  \bS_\text{taper}^{-1} \bD^\T \bV_\text{taper}^{-1} \bV \ \bV_\text{taper}^{-1} \bD \bS_\text{taper}^{-1}
\eeq
where $\bS_\text{taper} = \bD^\T \bV_\text{taper}^{-1} \bD$. Note that it is not
required to invert the non-sparse covariance matrix $\bV$ in order to compute \eqref{eq:tapercov}.

\section{ASYMPTOTIC THEORY}\label{sec:asymptotics}

Let $W_n \subset \R^2$ be an increasing sequence of observation windows in $\R^2$. Following Section~\ref{sec:gandtilderho} we assume that the true pair correlation function is given by a parametric model
$g(\br)=g(\br;\bpsi)$ for some unknown parameter vector $\bpsi\in \R^{q}$. Let $\btheta=(\bbt,\bpsi) \in \R^{p+q}$. We denote
the true value of $\btheta$ by $\btheta^*=(\bbt^*,\bpsi^*)$. In what
follows, $\EE$ and $\Var$ denote expectation and variance under the distribution
corresponding to $\btheta^*$.

Introducing the dependence on $n$ and $\btheta$  in the notation from
Section~\ref{sec:gee}, we have
\[
  \bphi_{n,\btheta}(\bu,\bbt) = \Big[ (\bI + \bT_{n,\btheta})^{-1}
   \frac{\blambda'(\cdot;\bbt)}{\lambda(\cdot;\bbt)} \Big](\bu), \quad (\bT_{n,\btheta} \bff)(\bu)
  = \int_{W_n} t_\btheta(\bu,\bv) \bff(\bv) \dd \bv
\]
and
\[
   t_\btheta(\bu,\bv) = \lambda(\bv;\bbt) \big[ g(\bu - \bv;\bpsi) - 1 \big].
\]
Following Section~\ref{sec:gandtilderho} we replace $\btheta$ in the kernel
$t_\btheta$ by a preliminary estimate $\tilde \btheta_n = (\tilde \bbt_n,\tilde \bpsi_n)$.
The estimating function \eqref{eq:limitgee} then becomes $\be_{n,\tilde \btheta_n}(\bbt)$
where
\[
   \be_{n,\btheta}(\bbt) = \sum_{\bu \in X \cap W_n} \bphi_{n,\btheta}(\bu,\bbt)
    - \int_{W_n} \bphi_{n,\btheta}(\bu,\bbt)\lambda(\bu;\bbt) \dd \bu.
\]
Let $\hat\bbt_n$ denote the estimator obtained by solving $\be_{n,\tilde \btheta_n}(\bbt) = 0$.
Further, define
\[
   \bar \bsig_n = |W_n|^{-1} \Var \be_{n,\btheta^*}(\bbt^*), \;
   \bJ_{n,\btheta}(\bbt) = -\frac{\dd}{\dd \bbt^\T} \be_{n,\btheta}(\bbt)
   \, \text{ and } \, \bar \bS_{n,\btheta}(\bbt) = |W_n|^{-1} \EE \bJ_{n,\btheta}(\bbt).
\]
Note that $\bar \bsig_n$ and $\bar \bS_{n,\btheta}(\bbt)$ are
`averaged' versions of $\bsig_n = \Var \be_{n,\btheta^*}(\bbt^*)$ and
$\bS_{n,\btheta}(\bbt) = \EE \bJ_{n,\btheta}(\bbt)$.

In Appendix~\ref{sec:existence} we verify the existence of a $|W_n|^{1/2}$ consistent
sequence of solutions $\hat \bbt_n$, i.e., $|W_n|^{1/2}(\hat \bbt_n - \bbt^*)$
is bounded in probability. We further show in Appendix~\ref{sec:asympnorm} that
$|W_n|^{-1/2} \be_{n,\tilde \btheta_n}(\bbt^*)\bar \bsig_n^{-1/2}$ is asymptotically
standard normal. The conditions needed for these results are listed in
Appendix~\ref{sec:conditions}. It then follows by a Taylor series expansion,
\[
   |W_n|^{-1/2} \be_{n,\tilde \btheta_n}(\bbt^*) \bar \bsig_n^{-1/2} =
   |W_n|^{1/2}(\hat \bbt_n - \bbt^*)\frac{\bJ_{n,\tilde \btheta_n}(\mathbf{b}_n)}{|W_n|} \bar \bsig_n^{-1/2}
\]
for some $\mathbf{b}_n\in\R^{p}$ satisfying
$\|\mathbf{b}_n - \bbt^*\| \leq \| \hat \bbt_n - \bbt^* \|$,
and R\ref{r:jnjn} and R\ref{r:jnin} in Appendix~\ref{sec:existence} that
\[
   |W_n|^{1/2}(\hat \bbt_n - \bbt^*) \bar \bS_{n,\btheta^*}(\bbt^*) \bar \bsig_n^{-1/2} \rightarrow N_p(0, \bI).
\]
Hence, for a fixed $n$ and since $\bar \bsig_n = \bar \bS_{n,\btheta^*}(\bbt^*)$ by \eqref{eq:SandSigma}, $\hat \bbt_n$ is approximately normal with mean
$\bbt^*$ and covariance matrix estimated by $|W_n|^{-1} \bar \bS^{-1}_{n,(\tilde \bpsi_n,\hat \bbt_n)} (\hat \bbt_n)$.


\section{SIMULATION STUDY AND DATA EXAMPLE}\label{sec:simulations_dataexamples}

To examine the performance of our optimal intensity estimator relative to composite likelihood and weighted composite likelihood, we carry out a
simulation study under the \cite{guan:shen:10} setting. We use the quasi-likelihood
implementation of our estimator as described in
Sections~\ref{sec:ql}-\ref{sec:tapering} and hence use the term
quasi-likelihood for our approach. We refrain from a comparison with
maximum likelihood estimation due to the lack of a computationally
feasible implementation of this method. In addition to the simulation
study we demonstrate the practical usefulness of our method and
discuss computational issues in a tropical rain forest data example.

\subsection{Simulation Study}\label{sec:simstudy}

In the simulation study, following \cite{guan:shen:10}, realizations of Cox
processes are generated on a square window $W$. Each simulation involves first
the generation of  a zero-mean Gaussian random field ${\bf Z}=\{Z(\bu)\}_{\bu \in W}$ with
exponential covariance function $c(\bu)=\exp(-\|\bu\|/0.1)$ and then
the generation of an inhomogeneous Thomas process given ${\bf Z}$ with intensity function
$\lambda(\bu;\bbt) = \exp\big[\beta_{0} + \beta_{1} Z(\bu)\big]$ and clustering parameter
$\bpsi=(\kp,\omega)$, cf.\ \eqref{eq:thomaspcf} in Appendix~\ref{sec:neumanncondition}. For each simulation $\bbt=(\bt_0,\bt_1)$
is estimated using composite likelihood (CL), weighted composite likelihood (WCL),
and quasi-likelihood (QL). The clustering parameter $\bpsi$ is estimated using minimum
contrast estimation based on the $K$-function  \cite[e.g.\ Section~10.1 in][]{moeller:waagepetersen:04}.

The simulation window is either $W=[0,1]^2$ or $W=[0,2]^2$. The mean square error (MSE)
of the CL, WCL and QL estimates is computed using 1000 simulations for each combination
of different clustering levels (i.e., different expected numbers of clusters $\kp^*=100$ or
$200$ and different cluster radii $\omega^*=0.02$ or $0.04$), inhomogeneity levels
($\beta_{1}^*=0.5$ or $1$), and expected number of points ($400$  in the
case of $W=[0, 1]^2$ and 1600  in the case of $W = [0,2]^2$). The integral terms
in the CL, WCL and QL estimating equations are approximated using a $50 \times 50$ grid
for $W = [0,1]^2$ and a $100 \times 100$ grid for $W = [0,2]^2$. Tapering for QL is
carried out as described in Section~\ref{sec:tapering} using $d_\text{taper}$ obtained
with $\epsilon=0.01$ for each estimated pair correlation function $g(\cdot;\hat \bpsi)$.
For WCL we use $A(\bu) \approx K(d_\text{taper}; \hat \bpsi)- \pi d_\text{taper}^2$ where
\[
   K(t;\bpsi)=\int_{\|\br\|\leq t}g(\br;\bpsi) \dd \br.
\]

Table~1 shows the reduction in MSE for the WCL and
 QL estimators relative to the CL estimator. The reductions show that one can obtain more
efficient estimates of the intensity function by taking into account the correlation
structure of the process. As expected from the theoretical results,
the QL estimator has superior performance compared with both the CL and the WCL estimators in all cases. The improvement over the CL estimator is especially substantial in the more clustered (corresponding to small $\kp^*$ and $\omega^*$) and more
inhomogeneous (corresponding to $\bt_1^*=1$) cases where the largest reduction is 68.5\%. As we alluded in Section 3.2, the performance of the WCL estimator may rely on the validity of the approximation \eqref{eq:wclapprox}. In case of a longer dependence range, the approximation is expected to be less accurate and this explains the large drop in the efficiency of the WCL estimator relative to the CL estimator when $\omega^*$ increases from 0.02 to 0.04. In particular, the WCL estimator does not appear to perform any better than the CL estimator when $\bpsi^*=(200,0.04)$. In contrast, the QL estimator still gives significant reductions in MSE of size 10-26\% depending on the value of $\bt_1^*$ and $W$.
\begin{table}
\caption{Reduction (\%) in MSE (summed for $\bt_0$ and $\bt_1$)
for WCL and QL relative to CL.}\label{tab:simstudy}
\centering
\begin{tabular}{c|cccc|cccc}
  & \multicolumn{4}{c|}{$W = [0,1]^2$} & \multicolumn{4}{c}{$W = [0,2]^2$} \\\hline
  \multirow{2}{*}{$\bpsi^*=(\kappa^*, \omega^*)$}
  & \multicolumn{2}{c}{$\beta_{1}^* = 0.5$} & \multicolumn{2}{c|}{$\beta_{1}^* = 1.0$}
  & \multicolumn{2}{c}{$\beta_{1}^* = 0.5$} & \multicolumn{2}{c}{$\beta_{1}^* = 1.0$}  \\
  & WCL & QL & WCL & QL & WCL & QL & WCL & QL \\\hline
(100, 0.02) & 15.6  & 35.9 & 41.4  & 59.3 & 17.2 & 39.7 & 52.2 & 68.5 \\
(100, 0.04) & 1.5   & 34.4 & 14.2  & 42.2 & 11.9  & 38.9 & 13.6 & 55.1 \\
(200, 0.02) & 4.9   & 15.4 & 20.2  & 34.0 & 8.6 & 19.9 & 26.3 & 40.0 \\
(200, 0.04) & -3.5  & 16.5 & 3.0  & 26.2 & 2.0 & 10.3 & -7.5  & 18.0
\end{tabular}
\end{table}

\subsection{Data Example}\label{sec:dataex}

A fundamental problem in biological research is to understand the very high biodiversity in tropical rain forests. One explanation is the niche assembly hypothesis, which states that different species coexist by adapting to different environmental niches. Data available for studying this hypothesis consist of point patterns of locations of trees as well as observations of environmental covariates.
\begin{figure}[!htb]
\centering
\includegraphics[width=9cm, height=4.5cm]{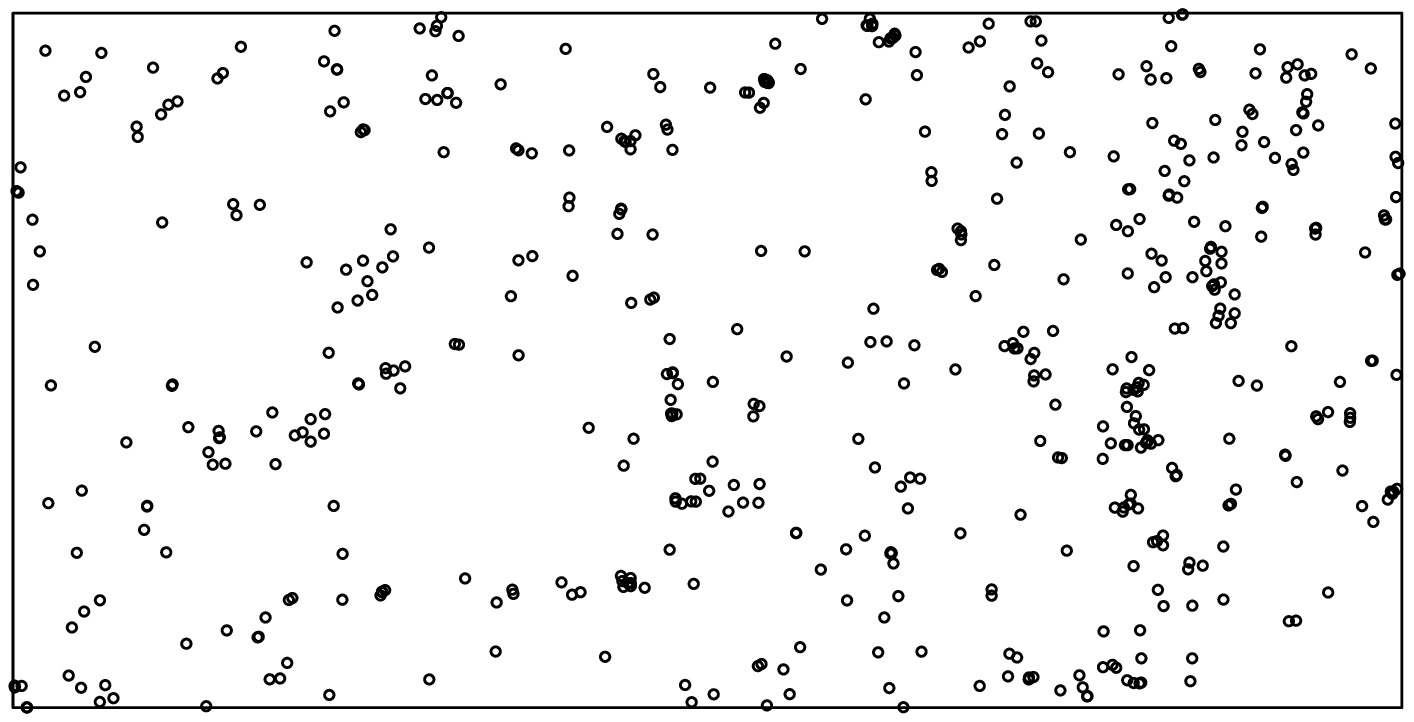}\\
\includegraphics[width=9cm, height=4.5cm]{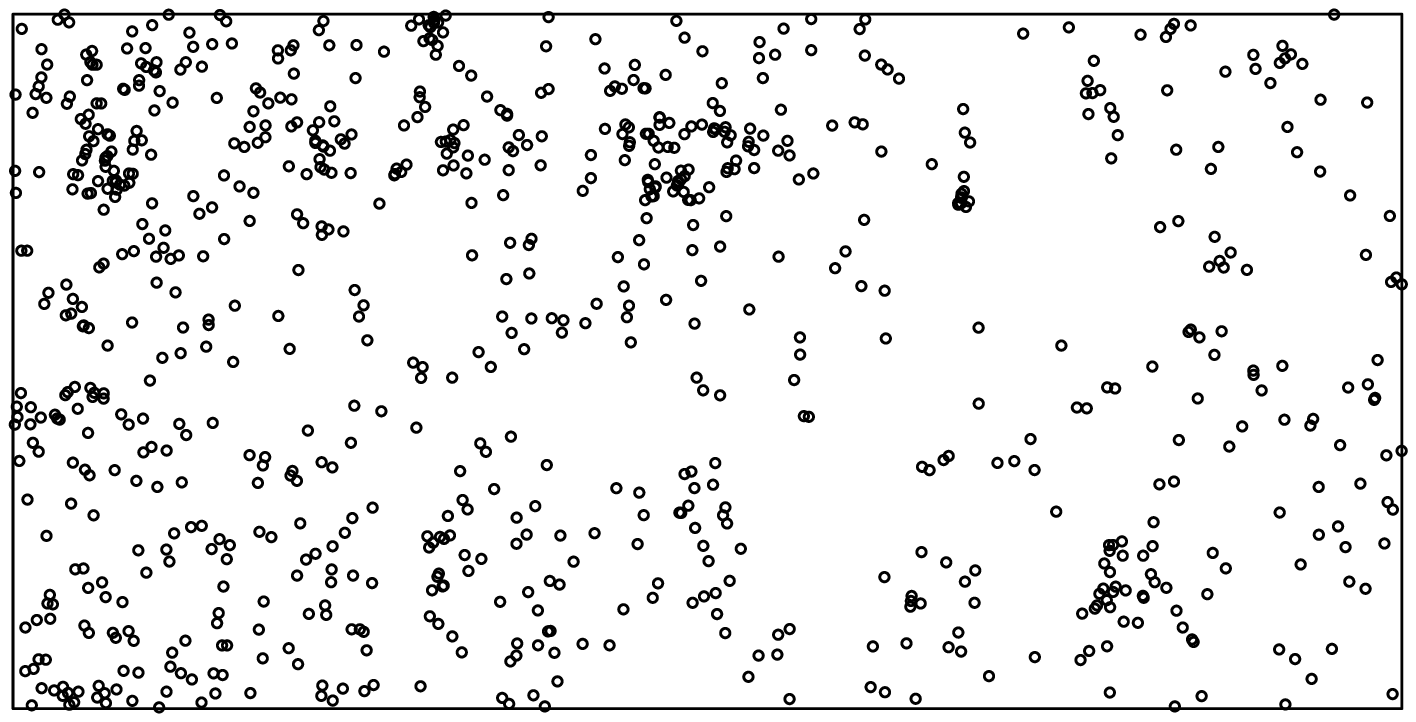}\\
\includegraphics[width=9cm, height=4.5cm]{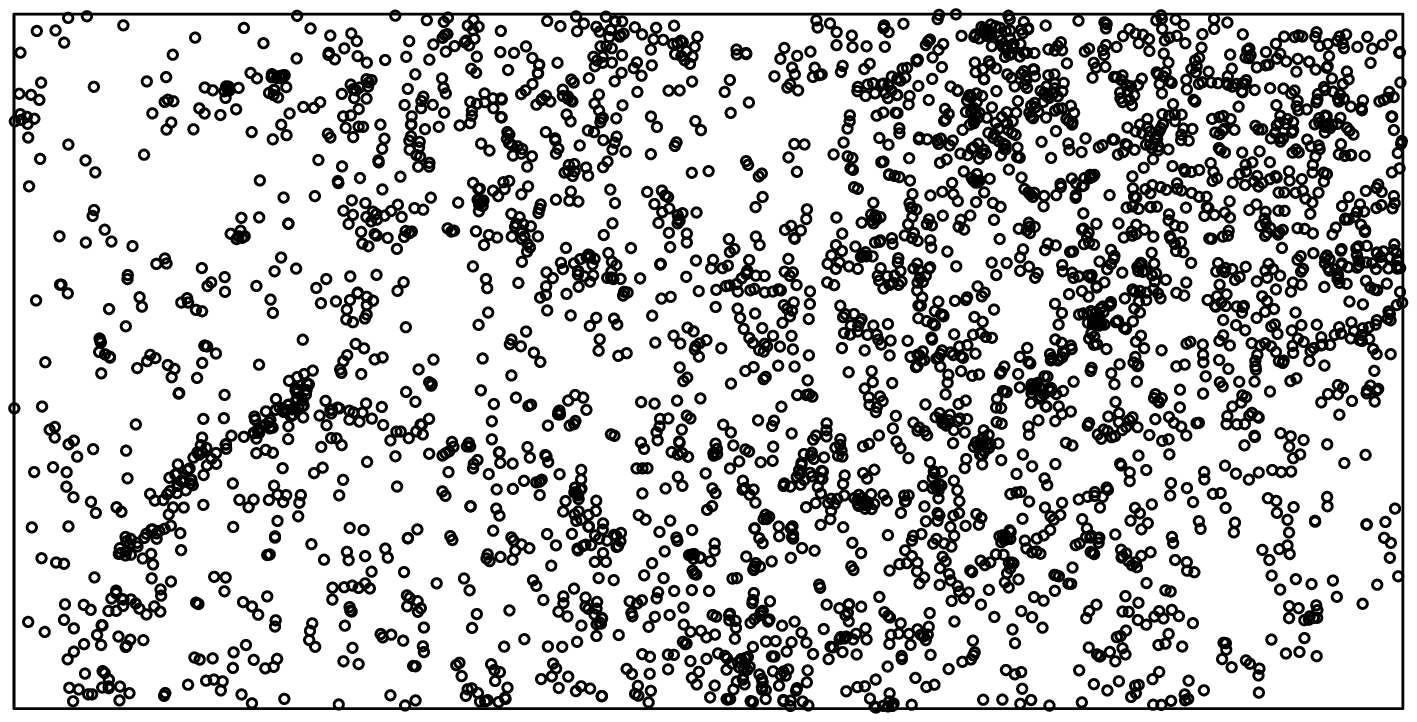}\\
\includegraphics[width=9cm, height=4.5cm]{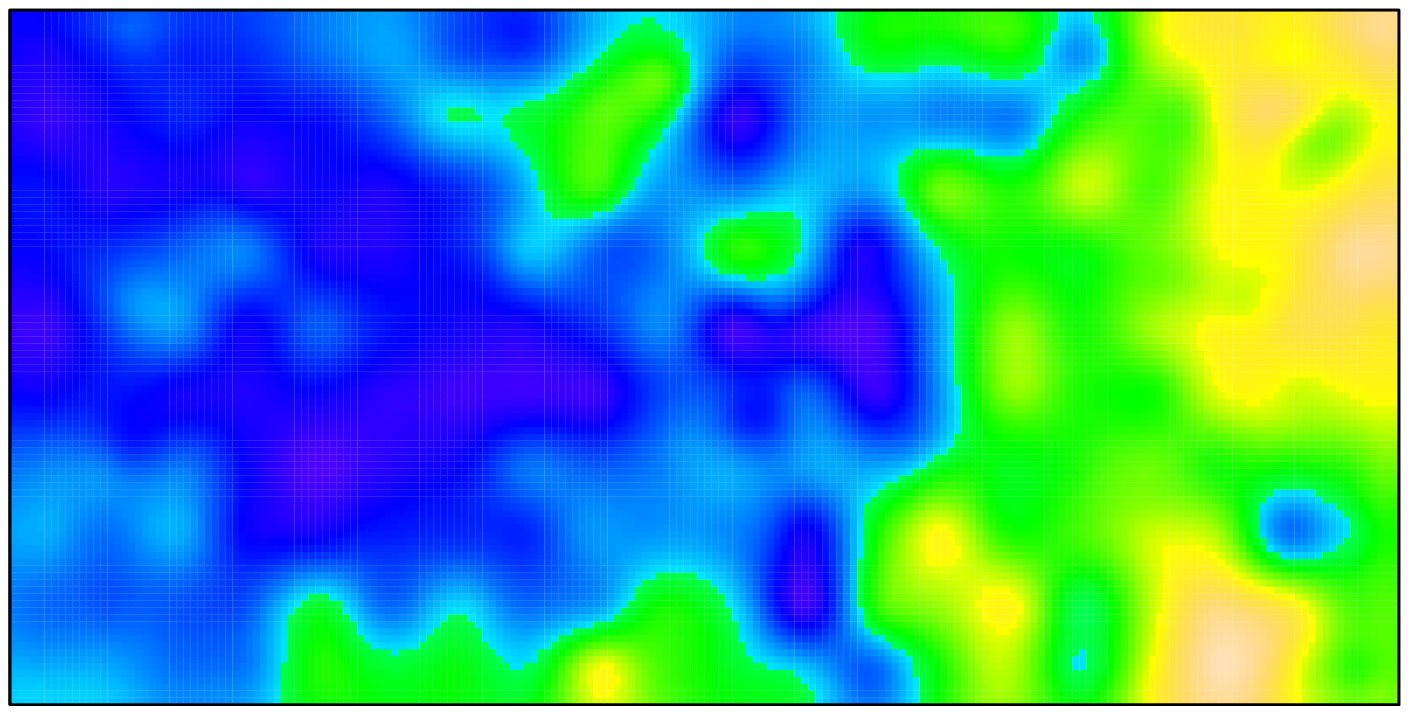}
\caption{Locations of Acalypha, Lonchocarpus, and Capparis trees and image of interpolated potassium content in the surface soil (from top to bottom).}\label{fig:data}
\end{figure}
Figure~\ref{fig:data} shows the spatial locations of three tree species,
{\em Acalypha diversifolia} (528 trees),
{\em Lonchocarpus heptaphyllus} (836 trees) and
{\em Capparis frondosa} (3299 trees), in a $1000m \times 500m$ observation window
on Barro Colorado Island \cite[][]{condit:hubbell:foster:96,condit:98,hubbell:foster:83}. Also one example of an environmental variable (potassium content in the soil) is shown.

In order to study the niche assembly hypothesis we use our quasi-likelihood method to fit log-linear regression models for the intensity functions depending on environmental variables. In addition to soil potassium content (\texttt{K}, divided by 1000), we consider nine other covariates for the intensity functions: pH, elevation (\texttt{dem}), slope gradient (\texttt{grad}),
multi-resolution index of valley bottom flatness (\texttt{mrvbf}), incoming mean solar radiation (\texttt{solar}), topographic wetness index (\texttt{twi}) as well as soil contents of copper (\texttt{Cu}), mineralized  nitrogen (\texttt{Nmin}) and phosphorus (\texttt{P}). The quasi-likelihood  estimation was implemented as in the simulation study using a $100 \times 50$ grid for the numerical quadrature and tapering tuning parameter $\epsilon=0.01$.

For each species we initially fit the following pair correlation functions of normal
variance mixture type \cite[][]{jalilian:guan:waagepetersen:10}:
$$
   g(\br;\bpsi)= 1 + c(\br;\bpsi), \quad \br \in \R^2,
$$
where the covariance function $c(\br;\bpsi)$ is either Gaussian
\[
   c(\br;(\sigma^2,\alpha)) = \sigma^2 \exp\big[-(\|\br\|/\af)^2 \big],
\]
Mat{\' e}rn ($K_\nu$ is the modified Bessel function of the second kind)
\[
   c(\br;(\sigma^2,\af,\nu)) = \sigma^2 \frac{(\|\br\|/\alpha)^{\nu} K_{\nu}(\|\br\|/\alpha)}{2^{\nu-1} \Gamma(\nu)},
\]
or Cauchy
\[
   c(\br;(\sigma^2,\af)) = \sigma^2 \big[ 1 +(\|\br\|/\af)^2 \big]^{-3/2}.
\]
These covariance functions represent very different tail behavior ranging from light (Gaussian), exponential (Mat{\' e}rn), to heavy tails (Cauchy).
The pair correlation function obtained with the Gaussian covariance function is
just a re-parametrization of the Thomas process pair correlation function \eqref{eq:thomaspcf}. For the
Mat{\'e}rn covariance we consider three different values of the shape parameter
$\nu=0.25, 0.5$ and $1$. With $\nu=0.5$ the exponential model $c[\br;(\sigma^2,\af,0.5)]=\sigma^2 \exp(-\|\br\|/\af)$
is obtained while $\nu=0.25$ and $1$ yields respectively a log convex and a log
concave covariance function.

\begin{figure}
\centering
\begin{tabular}{ccc}
\includegraphics[width=0.33\textwidth]{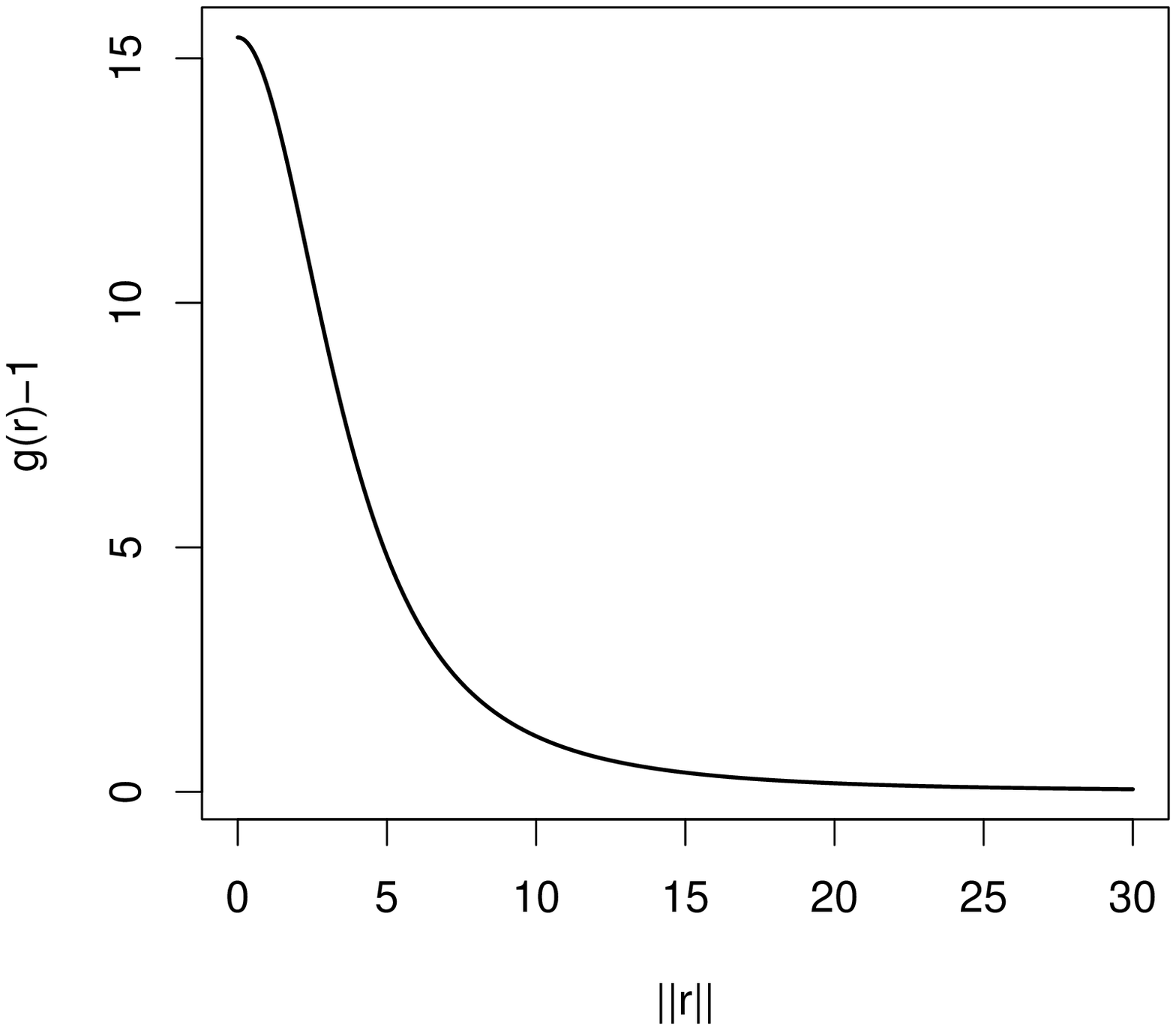} &
\includegraphics[width=0.33\textwidth]{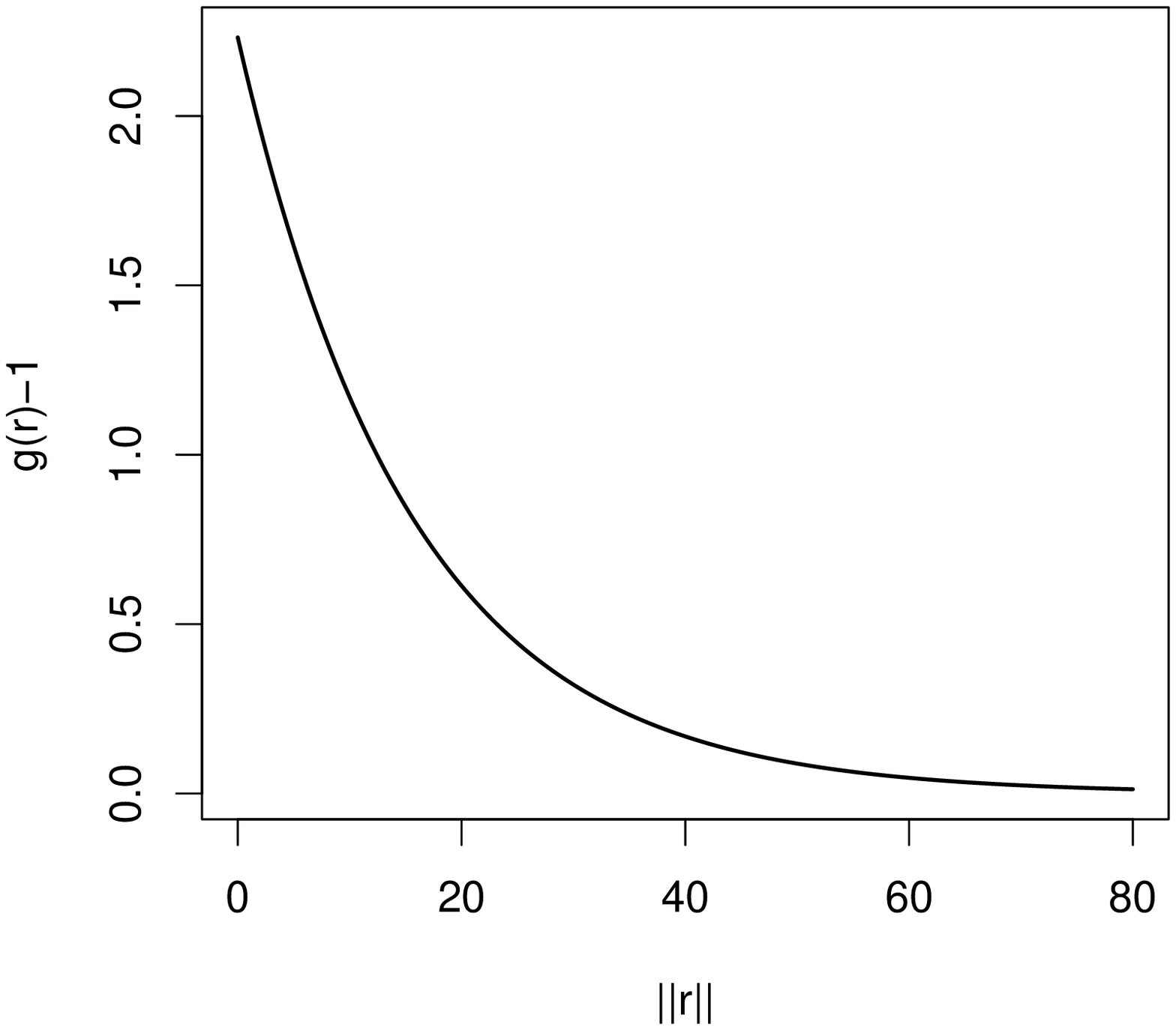} &
\includegraphics[width=0.33\textwidth]{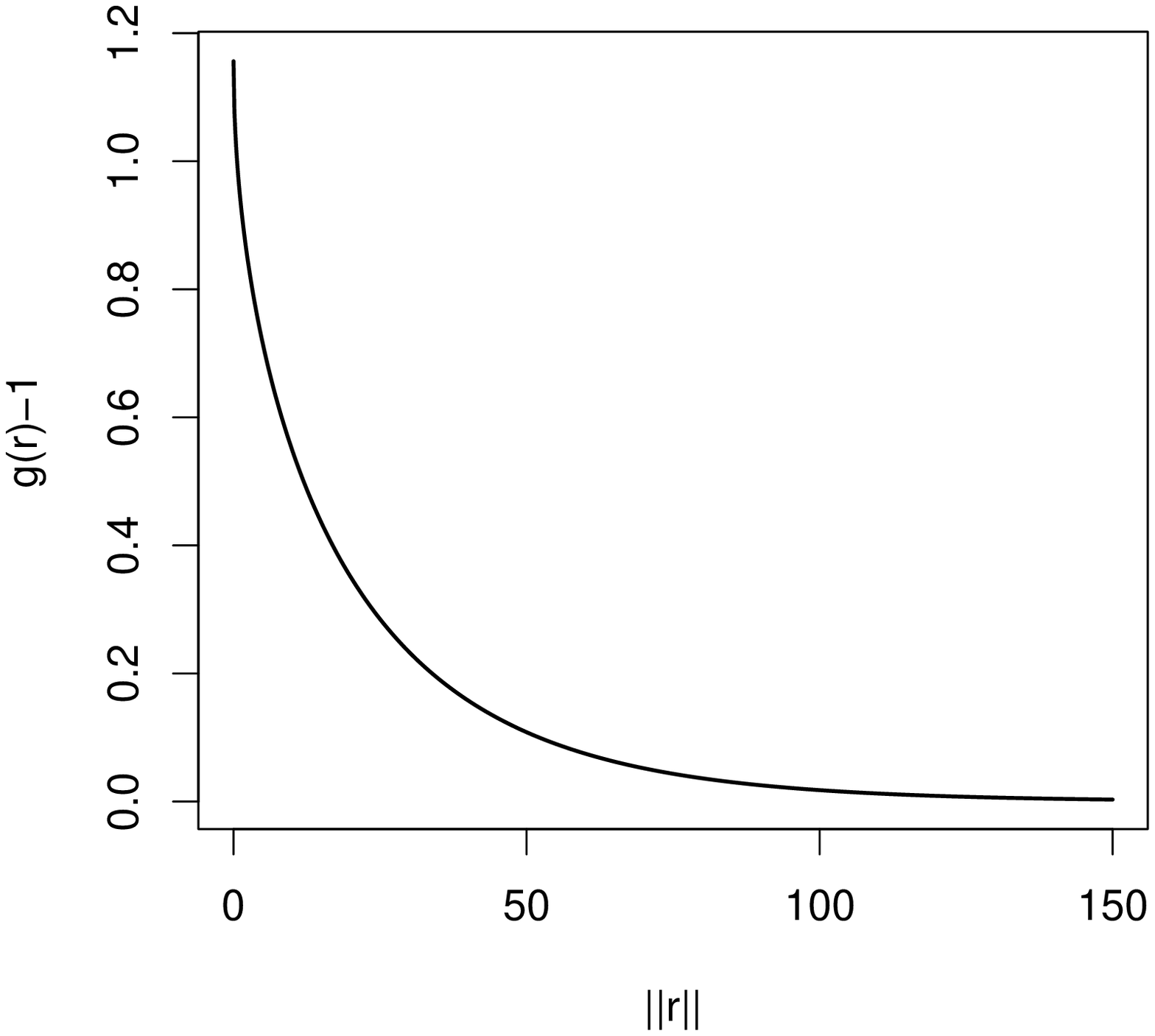}
\end{tabular}
\caption{Best fitting covariance functions $c(\cdot;\hat
  \bpsi)=g(\cdot;\hat \bpsi)-1$ for Acalypha (left), Lonchocarpus
  (middle), and Capparis (right).}\label{fig:covcompare}
\end{figure}

Figure~\ref{fig:covcompare} shows $c(\cdot;\hat \bpsi)=g(\cdot;\hat \bpsi)-1$ for the best fitting (in terms of the minimum contrast criterion for the corresponding $K$-function) pair correlation functions:  Cauchy for Acalypha ($\hat\bpsi=(15.4,2.3)$), Mat{\' e}rn ($\hat\bpsi=(2.2,15.5,0.5$)) for Lonchocarpus and Mat{\' e}rn ($\hat \bpsi=(1.2,30.2,0.25)$) for Capparis.
The tapering distances corresponding to $\epsilon=0.01$ are
respectively 20.9, 71.3 and 112.2 for the three species. Hence Capparis is the computationally most challenging case.

Backward model selection with significance level 5\% was carried out
for each species. According to the quasi-likelihood results, potassium
(\texttt{K}) is a significant covariate at the 5\% level for Acalypha,
mineralized nitrogen (\texttt{Nmin}) and phosphorous (\texttt{P}) are
significant for Lonchocarpus while elevation (\texttt{dem}), gradient
(\texttt{grad}) and potassium are  significant for Capparis. The fitted linear predictors with estimated standard errors in parenthesis are respectively -6.9+4.4\texttt{K} (0.085,1.2), -6.5-0.028\texttt{Nmin}-0.15\texttt{P} (0.088,0.0069,0.055) and -5.1+0.020\texttt{dem}-2.3\texttt{grad}+3.9\texttt{K} (0.078,0.0090,0.98,1.0).

The computing time for the QL estimation depends both on the grid
used for the numerical quadrature and the
tapering tuning parameter $\epsilon$. We also tried out a $150 \times
75$ grid and $\epsilon=0.05$ and $0.02$ for the QL fitting of the
final models. Parameter estimates and parameter estimation computing
time (system plus CPU time on a 2.90 GHz lap
top) for all combinations of grid sizes, $\epsilon$ and species are
shown in Table~2. The computing time for
the parameter estimation depends much on both grid sizes, $\epsilon$
and species (i.e.\ range of spatial dependence). Computing time including
computation of standard errors is shown in Table~3, together with the computed standard errors for the parameter
estimates in Table~2.
The computing time with computation of
standard errors is less sensitive to $\epsilon$ and species since in this
case the main computational burden arises from the non-sparse matrix in
\eqref{eq:tapercov}. For the $100 \times 50$ grid and
$\epsilon=0.01$, the maximal computing time of 29.1 seconds (including computation of standard errors) occurs for
Capparis. In contrast to large variations in the computing time, the parameter estimates
and estimated standard errors for each species are very stable across the combinations of grid sizes and tapering parameter $\epsilon$.
\begin{table}
\caption{Computing times (T) in seconds (without computation of
   standard errors) and QL parameter estimates for different
combinations of grid
size and tapering.}\label{tbl:computingtimespar}
\centering
\begin{tabular}{ll|ll|ll|ll}
     \multicolumn{2}{c|}{}         & \multicolumn{2}{c|}{Acalypha} & \multicolumn{2}{c|}{Lonchocarpus} & \multicolumn{2}{c}{Capparis} \\
Grid & $\epsilon$ & T & estm.\  & T & estm.\ & T & estm.\ \\ \hline
\multirow{3}{*}{$100\!\! \times\!\! 50$}
& 0.05 & 0.3 &  -6.9 4.4 & 1.1 & -6.5 -0.028 -0.16 & 2.4 & -5.1 0.021 -2.4 4.2  \\
&  0.01  & 0.4 & -6.9 4.4&2.6 & -6.5 -0.028 -0.15 & 7.5 &  -5.1 0.020 -2.3 3.9\\
&  .002  & 0.6  & -6.9 4.4 &4.4  & -6.5 -0.028 -0.15 & 12.7 & -5.1 0.020 -2.3 3.8 \\ \hline
\multirow{3}{*}{$150\!\! \times\!\! 75$}
&  0.05  & 0.5 & -6.9 4.3& 8.5 &-6.5 -0.028 -0.16 & 34.9 & -5.1  0.021 -2.3  4.1 \\
&  0.01  & 1.8 &-6.9  4.3&23.7 &-6.5 -0.028 -0.15 & 80.4 & -5.1  0.020 -2.2 3.8\\
&  .002  & 5.3  &-6.9 4.3 &41.6 &-6.5 -0.028 -0.15 &163.6 & -5.1 0.020 -2.2 3.8\\
\end{tabular}
\end{table}
\begin{table}
\caption{Computing times (T) in seconds (including computation of
standard errors) and estimated standard errors of QL parameter estimates for different
combinations of grid
   size and tapering}\label{tbl:ssd}
\centering
\begin{tabular}{ll|ll|ll|ll}
     \multicolumn{2}{c|}{}         & \multicolumn{2}{c|}{Acalypha} & \multicolumn{2}{c|}{Lonchocarpus} & \multicolumn{2}{c}{Capparis} \\
Grid & $\epsilon$ & T & sd.\  & T & sd.\ & T & sd.\ \\ \hline
\multirow{3}{*}{$100\!\! \times\!\! 50$}
& 0.05 & 12.1 &0.085 1.2 & 22.4 & 0.088 0.0069 0.055  & 24.7 &    0.078 0.0091 0.98 1.1\\
&  0.01  & 12.0 &0.085 1.2& 24.0 & 0.088 0.0069 0.055 & 29.1 &  0.078 0.0090 0.98 1.0\\
&  .002  & 12.1  &0.085 1.2&25.9 & 0.088 0.0069 0.055 & 34.3 &  0.078 0.0090 0.98 1.0\\ \hline
\multirow{3}{*}{$150\!\! \times\!\! 75$}
&  0.05  & 59.4 &0.079 1.1& 187.2& 0.087 0.0069 0.055  & 223.4 & 0.078 0.0090 0.96 1.0 \\
&  0.01  & 58.9 &0.079 1.1&204.6 & 0.087 0.0069 0.055 & 255.2 & 0.078 0.0089 0.96 1.0\\
&  .002  & 63.6 &0.079 1.1 & 226.5 & 0.087 0.0069 0.055 & 300.9 & 0.078 0.0089 0.96 1.0 \\
\end{tabular}
\end{table}

\section{DISCUSSION}

In contrast to maximum likelihood estimation our quasi-likelihood estimation method only requires the specification
of the intensity function and a pair correlation function. Moreover, the estimation
of the regression parameters can be expected to be quite robust toward misspecification of the pair
correlation function since the resulting estimating equation is unbiased for any
choice of pair correlation function. In the data example we considered pair
correlation functions obtained from covariance functions of normal variance mixture
type. Alternatively one might consider pair correlation functions of the log Gaussian
Cox process type \citep{moeller:syversveen:waagepetersen:98}, i.e., $g(\br) = \exp\big[  c(\br) \big]$, where $c(\cdot)$ is an arbitrary covariance function.

If a log Gaussian Cox process is deemed appropriate, a computationally feasible
alternative to our approach is to use the method of integrated nested Laplace
approximation \cite[INLA,][]{rue:martino:chopin:09,illian:etal:12} to implement Bayesian inference.
However, in order to apply INLA it is required that the Gaussian field can be
approximated well by a Gaussian Markov random field and this can limit the choice
of covariance function. For example, the accurate Gaussian Markov random field
approximations in \cite{lindgren:rue:lindstrom:11} of Gaussian fields with
Mat{\'e}rn covariance functions are restricted to integer $\nu$ in the planar case.
In contrast, our approach is not subject to such limitations and can also be applied
to non-log Gaussian Cox processes.

We finally note that for the Nystr{\"o}m approximate solution of the Fredholm equation we used the simplest possible quadrature scheme given by a Riemann sum for a fine grid. This entails a minimum of assumptions regarding the integrand but at the expense of a typically high-dimensional covariance matrix $\bV$. There may hence be scope for further development considering more sophisticated numerical quadrature schemes.
\\[\bsl]
{\bf Acknowledgments}\\[\bsl]
Abdollah Jalilian and Rasmus Waagepetersen's research was supported by the Danish Natural Science Research Council, grant 09-072331 `Point process modeling and statistical inference',  Danish Council for Independent Research | Natural Sciences, Grant 12-124675, `Mathematical and Statistical Analysis of Spatial Data', and by Centre for Stochastic Geometry and Advanced Bioimaging, funded by a grant from the Villum Foundation. Yongtao Guan's research was supported by NSF grant DMS-0845368, by NIH grant 1R01DA029081-01A1 and by the VELUX Visiting Professor Programmme.

The BCI forest dynamics research project was made possible by National
Science Foundation grants to Stephen P. Hubbell: DEB-0640386,
DEB-0425651, DEB-0346488, DEB-0129874, DEB-00753102, DEB-9909347,
DEB-9615226, DEB-9615226, DEB-9405933, DEB-9221033, DEB-9100058,
DEB-8906869, DEB-8605042, DEB-8206992, DEB-7922197, support from the
Center for Tropical Forest Science, the Smithsonian Tropical Research
Institute, the John D. and Catherine T. MacArthur Foundation, the
Mellon Foundation, the Celera Foundation, and numerous private
individuals, and through the hard work of over 100 people from 10
countries over the past two decades. The plot project is part of the Center for Tropical Forest Science, a global network of large-scale demographic tree plots.

The BCI soils data set were collected and analyzed by J.\ Dalling,
R.\ John, K.\ Harms, R.\ Stallard and J.\ Yavitt  with support from NSF DEB021104, 021115, 0212284, 0212818 and OISE 0314581, STRI and CTFS. Paolo Segre and Juan Di Trani provided assistance in the field. The covariates \texttt{dem}, \texttt{grad}, \texttt{mrvbf}, \texttt{solar} and \texttt{twi} were computed in SAGA GIS by Tomislav Hengl (\texttt{http://spatial-analyst.net/}).

\bibliographystyle{Chicago}
\bibliography{masterbib}

\appendix

\makeatletter   
\renewcommand{\@seccntformat}[1]{APPENDIX~{\csname the#1\endcsname}.\hspace*{1em}}
\makeatother

\section{Condition for optimality}\label{app:optimality}

To show that \eqref{eq:sufficient} implies non-negative definiteness of \eqref{eq:godambediff}, let $\hat \be_\bphi(\bbt) = \be_\bff(\bbt) \bsig_\bff^{-1} \bsig_{\bff\bphi}$
be the optimal linear predictor of $\be_\bphi(\bbt)$ given $\be_\bff(\bbt)$.
Then
\[
  \Var[\hat \be_\bphi(\bbt) - \be_\bphi(\bbt)]= \bsig_\bphi
  - \bsig_{\bphi \bff}\bsig^{-1}_\bff \bsig_{\bff \bphi}
\]
is non-negative definite whereby
\[
  \bS_\bphi \bsig_\bphi^{-1} \bS_\bphi -
  \bS_\bphi \bsig_\bphi^{-1} \bsig_{\bphi \bff}
  \bsig_\bff^{-1} \bsig_{\bff \bphi} \bsig_{\bphi}^{-1} \bS_\bphi
\]
is non-negative definite too. Hence, \eqref{eq:godambediff} is non-negative definite
provided
\[
   \bS_\bphi \bsig_\bphi^{-1} \bsig_{\bphi \bff} = \bS_\bff
\]
which follows from \eqref{eq:sufficient}
(in particular,  \eqref{eq:sufficient} implies $\bsig_\bphi = \bsig_{\bphi \bphi} = \bS_\bphi$).

\section{SOLUTION USING NEUMANN SERIES EXPANSION}\label{sec:neumanncondition}

Suppose that $\|\bT \|_{\text{op}} = \sup \{ \|\bT\bff\|_{\infty}/\|\bff\|_\infty
\: : \: \|\bff\|_\infty \neq 0\}<1$ where $\|\bff\|_\infty$ denotes the supremum
norm of a continuous function $\bff$ on $W$. Then we can obtain the solution $\bphi$
of \eqref{eq:fredholm} using a Neumann series expansion which may provide additional
insight on the properties of $\bphi$. More specifically,
\begin{equation}\label{eq:neumann}
   \bphi  = \sum_{k=0}^\infty (-\bT)^k \frac{\blambda'}{\lambda}.
\end{equation}
If the infinite sum in \eqref{eq:neumann} is truncated to the first term ($k=0$) then
\eqref{eq:limitgee} becomes the Poisson score.
Note that 
\[
   \|\bT\|_\infty \leq \sup_{\bu \in W} \int_W |t(\bu, \bv)| \dd \bv.
\]
Hence, a sufficient condition for the validity of the Neumann series expansion is
\begin{equation}\label{eq:suffcondneumann}
   \sup_{\bu \in W} \lambda(\bu;\bbt) \int_{\R^2} \big| g(\br) - 1\big| \dd \br < 1.
\end{equation}

Condition \eqref{eq:suffcondneumann} roughly requires that $g(\br) - 1$ does not
decrease too slowly to zero and/or that $\lambda$ is moderate. For example, suppose
that $g$ is the pair correlation function of a Thomas cluster process \cite[e.g.][Chapter~5]{moeller:waagepetersen:04},
\begin{equation}\label{eq:thomaspcf}
  g(\br) - 1 = \exp\big[-\|\br\|^2/ (4 \omega^2) \big]/(4 \pi  \omega^2  \kp), \;  \hbox{ for some }\kp,  \omega >0,
\end{equation}
where $\kp$ is the intensity of the parent process and $\omega$ is the normal dispersal parameter. Then,
\[
   \int_{\R^2} \big| g(\br) - 1 \big| \dd \br = \frac{1}{4 \pi  \kp   \omega^2}
     \int_{\R^2} \exp(- \frac {\| \br \|^2}{4\omega^2}) \dd \br = 1/\kp
\]
and \eqref{eq:suffcondneumann} is equivalent to $\sup_{\bu\in W}
\lambda(\bu;\bbt)< \kp.$
In this case, Condition \eqref{eq:suffcondneumann} can be quite restrictive. However, the Neumann series expansion is not essential
for our approach and we use it only for checking the conditions for asymptotic results; see Appendix~\ref{sec:conditions}.

\section{CONDITIONS AND LEMMAS}\label{sec:conditions}

To verify the existence of a $|W_n|^{1/2}$ consistent sequence of solutions
$\hat\bbt_n$, we assume that the following conditions are satisfied:
\begin{enumerate}
\renewcommand{\theenumi}{\arabic{enumi}}
\renewcommand{\labelenumi}{C\theenumi}
\item \label{c:rhobounded}
   $\lambda(\bu;\bbt) = \lambda(\bz(\bu)\bbt^\T)$ where $\lambda(\cdot)>0$
   is twice continuously differentiable and \\ $\sup_{\bu \in \R^2} \|\bz(\bu)\| < K_1$
   for some $K_1 < \infty$.
\item \label{c:paircorrelation}
   for some $0 < K_2 < \infty$,
   $\int_{\R^2} \big|g(\br;\bpsi^*) - 1 \big| \dd \br \leq K_2$.
\item \label{c:phibounded}
   $\bphi_{n,\btheta}(\bu,\bbt)$ is differentiable with
   respect to $\btheta$ and $\bbt$, and for $|\bphi_{n,\btheta}(\bu,\bbt)|$,
   $|\dd \bphi_{n,\btheta}(\bu,\bbt)/\dd \bbt|$ and $|\dd \bphi_{n,\btheta}(\bu,\bbt)/\dd \btheta|$,
   the supremum over $\bu \in \R^2, \bbt \in b(\bbt^*,K_3), \btheta \in b(\btheta^*,K_3)$
   is bounded for some $K_3 > 0$, where $b(\mathbf{x}, r)$ denotes the ball centered at
   $\mathbf{x}$ with radius $r>0$.
\item \label{c:tildetan}
   $|W_n|^{1/2}(\tilde \btheta_n - \btheta^*)$ is bounded in probability.
\item \label{c:Inposdef}
  $l=\liminf_n l_n > 0$, where for each $n$, $l_n$ denotes the minimal eigenvalue of
  \[
     \bar \bS_{n,\btheta^*}(\bbt^*) = |W_n|^{-1} \EE \bJ_{n,\btheta^*}(\bbt^*)
     = |W_n|^{-1}\int_{W_n} \bphi_{n,\btheta^*}(\bu)^\T \blambda'(\bu;\bbt^*) \dd \bu.
  \]
\end{enumerate}
Condition C\ref{c:rhobounded} and C\ref{c:paircorrelation} imply
L\ref{l:rhobounded} and L\ref{l:variancebounded} below.
\begin{enumerate}
\renewcommand{\theenumi}{\arabic{enumi}}
\renewcommand{\labelenumi}{L\theenumi}
\item \label{l:rhobounded}
   for $\lambda(\bu;\bbt)$, $\blambda'(\bu;\bbt)$ and $\blambda''(\bu;\bbt)$, the supremum over
   $\bu \in \R^2, \bbt \in b(\bbt^*,K_3), \btheta \in b(\btheta^*,K_3)$ is bounded.
\item \label{l:variancebounded}
   for a function $h:\R^2\to\R$,
  \[
     \Var \sum_{\bu \in X \cap W_n} h(\bu) \le |W_n|
     \big[ 1 + \sup_{\bu \in W_n} \lambda(\bu;\bbt^*) K_2 \big]
     \sup_{\bu \in W_n} h(\bu)^2 \sup_{\bu \in W_n}\lambda(\bu;\bbt^*).
  \]
In particular, $|W_n|^{-1}\Var \sum_{\bu \in X \cap W_n} h(\bu)$ is bounded
when $h$ is bounded.
\end{enumerate}

The condition C\ref{c:phibounded} is not so easy to verify in general due to the
abstract nature of the function $\bphi_{n,\btheta}$. However, it can be verified
e.g.\ assuming that $\bphi_{n,\btheta}$ can be expressed using the Neumann series.
Condition C\ref{c:tildetan} holds under conditions specified in \cite{waagepetersen:guan:09}
(including e.g.\ C\ref{c:rhobounded} and C\ref{c:paircorrelation}).
Condition C\ref{c:Inposdef} is not unreasonable since
\[
   \bar \bS_{n,\btheta^*}(\bbt^*) = |W_n|^{-1} \int_{W_n} \Big[ \frac{\blambda'(\bu;\bbt^*)}{\lambda(\bu;\bbt^*)^{1/2}} \Big]^\T
   \Big [ (\bI + \bT^s_{n,\btheta^*})^{-1} \frac{\blambda'(\cdot;\bbt^*)}{\lambda(\cdot;\bbt^*)^{1/2}} \Big ](\bu) \dd \bu
\]
and $(\bI + \bT^s_{n,\btheta^*})^{-1}$ is a positive operator (see Section~\ref{sec:nonnegeigen}).
Since $\bar \bsig_n = \bar \bS_{n,\btheta^*}(\bbt^*)$, C\ref{c:Inposdef} also implies
\begin{enumerate}
\renewcommand{\theenumi}{\arabic{enumi}}
\renewcommand{\labelenumi}{L\theenumi}
\addtocounter{enumi}{2}
\item \label{l:sigmain}
   $l=\liminf_n l_n > 0$ where for each $n$, $l_n$ denotes the  minimal eigenvalue
   of $\bar \bsig_n$.
\end{enumerate}

To prove the asymptotic normality of  $|W_n|^{-1/2}
\be_{n,\tilde\btheta_n}(\bbt^*) \bar \bsig_n^{-1/2}$,
we assume that the following additional conditions are satisfied:
\begin{enumerate}
\renewcommand{\theenumi}{\arabic{enumi}}
\renewcommand{\labelenumi}{N\theenumi}
\item \label{N:W_n}
   $W_n = nA$ where $A\subset (0,1]\times (0,1]$ is the interior of a simple closed
   curve with nonempty interior.

\item \label{N:mixing}
   $\sup_p{\alpha(p;k)\over p} = \hbox{O}(k^{-\epsilon})$ for some $\epsilon>2$,
   where $\alpha(p;k)$ is the strong mixing coefficient \citep{rosenblatt:56}. For
   each $p$ and $k$, the mixing condition measures the dependence between $X \cap E_1$
   and $X \cap E_2$ where $E_1$ and $E_2$ are arbitrary Borel subsets of $\R^2$ each of
   volume less than $p$ and at distance $k$ apart.

\item \label{N:cumulant}
   for some $K_4 < \infty$ and $k = 3,4$,
   \[
      \sup_{\bu_1\in\mathbb{R}^2} \int_{\R^2}\cdots\int_{\R^2}
      \big|Q_k(\bu_1,\cdots,\bu_k)\big| \dd \bu_2\cdots\dd \bu_k < K_4,
   \]
   where $Q_k$ is the $k$-th order cumulant density function of $X$ \cite[e.g.][]{guan:loh:07}.
\end{enumerate}
Conditions N1-N3 correspond to conditions (2), (3) and (6), respectively,
in \cite{guan:loh:07}. See this paper for a discussion of the conditions.

\section{EXISTENCE OF A $|W_{\lowercase{n}}|^{1/2}$ CONSISTENT $\hat \bbt_{\lowercase{n}}$}\label{sec:existence}

We use Theorem~2 and Remark~1 in \cite{waagepetersen:guan:09} to show the
existence of a $|W_n|^{1/2}$ consistent sequence of solutions $\hat\bbt_n$.
Let $\|\bA\|_M = \sup_{ij}|a_{ij}|$ for a matrix $\bA = [a_{ij}]_{ij}$.
With $\bV_n = |W_n|^{1/2} \bar \bsig_n^{1/2}$ we need to verify the following results:
\begin{enumerate}
\renewcommand{\theenumi}{\arabic{enumi}}
\renewcommand{\labelenumi}{R\theenumi}
\item \label{r:Vn}
   $\| \bV_n^{-1} \|_M \rightarrow 0.$
\item \label{r:jnjn}
   For any $d > 0$,
   \[
       \sup_{\bbt: \|(\bbt - \bbt^*)\bV_{n}\| \leq d} \| \bV_n^{-1}
       \big[\bJ_{n,\tilde \btheta_n}(\bbt) - \bJ_{n,\tilde \btheta_n}(\bbt^*) \big] \bV_n^{-1} \|_M
   \]
   converges to zero in probability.
\item \label{r:jnin}
   $\|\bJ_{n,\tilde \btheta_n}(\bbt^*)/|W_n| - \bar \bS_{n,\btheta^*}(\bbt^*)\|_M$
   converges to zero in probability.
\item \label{r:boundedinprob}
   $\be_{n,\tilde \btheta_n}(\bbt^*)\bV_n^{-1}$ is bounded in probability.
\item \label{r:posdef}
   $\liminf_n l_n>0$ where
   \[
      l_n = \inf_{\|\bx\| = 1} \bx \bar \bsig_n^{-1/2} \bar
      \bS_{n,\btheta^*}(\bbt^*) \bar \bsig_n^{-1/2} \bx^\T.
   \]
\end{enumerate}
We now demonstrate that R\ref{r:Vn}-R\ref{r:posdef} hold under
the conditions C\ref{c:rhobounded}-C\ref{c:Inposdef} listed in
Appendix~\ref{sec:conditions}. For each of the results below the required
conditions or previous results are indicated in square brackets.\\[\bsl]
R\ref{r:Vn} [C\ref{c:phibounded}, L\ref{l:rhobounded}-L\ref{l:sigmain}]:
By C\ref{c:phibounded}, L\ref{l:rhobounded} and L\ref{l:variancebounded}
the entries in $\bar \bsig_n$ are bounded from below and above. Moreover,
by L\ref{l:sigmain} the determinant of $\bar \bsig_n$ is bounded below by $l^p>0$.\\[\bsl]
R\ref{r:jnjn} [R\ref{r:Vn}, C\ref{c:phibounded}, L\ref{l:rhobounded}, L\ref{l:variancebounded}, C\ref{c:tildetan}]:
We show that
\[
   \sup_{(\btheta,\bbt):\|(\btheta - \btheta^*, \bbt - \bbt^*)|W_n|^{1/2}\| \leq d}
   \| |W_n|^{-1}\big[ \bJ_{n,\btheta}(\bbt) - \bJ_{n,\btheta^*}(\bbt^*) \big] \|_M
\]
converges to zero in probability. Note
\[
   |W_n|^{-1} \bJ_{n,\btheta}(\bbt)  = \bL_{n,\btheta}(\bbt) + \bM_{n,\btheta}(\bbt)
\]
where
\[
   \bL_{n,\btheta}(\bbt) = -\sum_{\bu \in X} \bff_{1,n,\btheta}(\bu,\bbt)
   \text{ and }
   \bM_{n,\btheta}(\bbt) = \int_{\R^2} \bff_{2,n,\btheta}(\bu,\bbt)
\]
with
\[
   \bff_{1,n,\btheta}(\bu,\bbt) = \frac{1[\bu \in W_n]}{|W_n|}
  \frac{\dd }{\dd \bbt^\T} \bphi_{n,\btheta}(\bu,\bbt)
\]
and
\[
   \bff_{2,n,\btheta}(\bu,\bbt)= \frac{1[\bu \in W_n]}{|W_n|}
   \big[ \lambda(\bu;\bbt) \frac{\dd }{\dd \bbt^\T} \bphi_{n,\btheta}(\bu,\bbt) +
   \blambda'(\bu;\bbt)^\T \bphi_{n,\btheta}(\bu,\bbt) \big].
\]
Define
\[
   h_{i,n}(\bu) = \sup_{(\btheta,\bbt): \|(\btheta - \btheta^*, \bbt - \bbt^*)|W_n|^{1/2}\| \leq d}
   |\bff_{i,n,\btheta}(\bu,\bbt) - \bff_{i,n,\btheta^*}(\bu,\bbt^*)|, \, i=1,2
\]
and note that $h_{i,n}(\bu)$ converge to zero as $n \rightarrow \infty$. Then
\[
   \sup_{(\btheta,\bbt):\|(\btheta - \btheta^*,\bbt - \bbt^*)|W_n|^{1/2}\| \leq d}
   |\bM_{n,\btheta}(\bbt) - \bM_{n,\btheta^*}(\bbt^*)| \leq \int_{\R^2} h_{1,n}(\bu) \dd \bu \]
where the right hand side converges to zero by dominated convergence. Moreover,
\begin{multline*}
  \sup_{(\btheta,\bbt):\|(\btheta - \btheta^*, \bbt - \bbt^*)|W_n|^{1/2}\| \leq d}
  \big|\bL_{n,\btheta}(\bbt) - \bL_{n,\btheta^*}(\bbt^*) \big|  \leq
  \sum_{\bu \in X} h_{2,n}(\bu) \leq \\
  \Big| \sum_{\bu \in X} h_{2,n}(\bu) - \EE \sum_{\bu \in X} h_{2,n}(\bu) \Big| +
  \Big| \EE \sum_{\bu \in X} h_{2,n}(\bu) \Big|.
\end{multline*}
The first term on the right hand side converges to zero in probability by Chebyshev's
inequality and the second term converges to zero by dominated convergence.
\\[\bsl]
R\ref{r:jnin} [R\ref{r:Vn}, L\ref{c:rhobounded}, L\ref{l:variancebounded}, C\ref{c:tildetan}]:
\begin{multline*}
  |W_n|^{-1}  \bJ_{n,\tilde \btheta_n}(\bbt^*) - \bar \bS_n(\bbt^*)  = \\
  |W_n|^{-1} \big[ \bJ_{n,\tilde \btheta_n}(\bbt^*) -
  \bJ_{n,\btheta^*}(\bbt^*) \big] + \big [
  |W_n|^{-1}  \bJ_{n,\btheta^*}(\bbt^*) - \bar \bS_n(\bbt^*) \big]
\end{multline*}
It follows from the proof of R\ref{r:jnjn} that the first term on the right hand
side converges to zero in probability. The last term converges to zero in
probability by Chebyshev's inequality.\\[\bsl]
R\ref{r:boundedinprob} [C\ref{c:phibounded}, L\ref{l:rhobounded}, L\ref{l:variancebounded}, C\ref{c:tildetan}]:
Since $\Var \be_{n,\btheta^*}(\bbt^*) \bV_n^{-1}$ is the identity matrix,
$\be_{n,\btheta^*}(\bbt^*)\bV_n^{-1}$ is bounded in probability by Chebyshev's
inequality. The result then follows by showing that\\
$|W_n|^{-1/2}\big[ \be_{n,\tilde \btheta_n}(\bbt^*) - \be_{n,\btheta^*}(\bbt^*) \big]$
converges to zero in probability. Let
\begin{multline*}
   \bff_n(\btheta) = |W_n|^{-1} \frac{\dd}{\dd \btheta^\T} \be_{n,\btheta}(\bbt^*) =\\
   |W_n|^{-1} \Big[ \sum_{\bu \in X \cap W_n} \frac{\dd}{\dd \btheta^\T } \bphi_{n,\btheta}(\bu,\bbt^*) -
    \int_{W_n} \lambda(\bu;\bbt^*) \frac{\dd}{\dd \btheta^\T } \bphi_{n,\btheta}(\bu,\bbt^*) \dd \bu \Big].
\end{multline*}
Then
\begin{equation*}
   |W_n|^{-1/2} \big[ \be_{n,\tilde \btheta_n}(\bbt^*) - \be_{n,\btheta^*}(\bbt^*) \big]
   = |W_n|^{1/2} (\tilde \btheta_n - \btheta^*) \bff_n(\mathbf{t}_n)
\end{equation*}
where $\|\mathbf{t}_n - \btheta^*\| \leq \|\tilde \btheta_n - \btheta^*\|$
and the factor $|W_n|^{1/2}(\tilde \btheta_n - \btheta^*)$ is bounded in probability.
Further,
\[
   \bff_n(\mathbf{t}_n) = \bff_n(\mathbf{t}_n) - \bff_n(\btheta^*) + \bff_n(\btheta^*)
\]
where $\bff_n(\btheta^*)$ converges to zero in probability by Chebyshev's inequality
and $\bff_n(\mathbf{t}_n) - \bff_n(\btheta^*)$ converges to zero in probability along
the lines of the proof of R\ref{r:jnjn}.\\[\bsl]
R\ref{r:posdef} [C\ref{c:Inposdef}, L\ref{l:sigmain}]: Follows directly from C\ref{c:Inposdef} and L\ref{l:sigmain}.

\section{ASYMPTOTIC NORMALITY OF $|W_{\lowercase{n}}|^{-1/2} \be_{\lowercase{n},\tilde \btheta_{\lowercase{n}}}(\bbt^*) \bsig_{\lowercase{n}}^{-1/2}$}\label{sec:asympnorm}

By the proof of R\ref{r:boundedinprob} it suffices to show that
$|W_n|^{-1/2}  \be_{n,\btheta^*}(\bbt^*) \bar \bsig_n^{-1/2}$ is asymptotically normal.
To do so we use the blocking technique used in \cite{guan:loh:07}. Specifically,
Condition N\ref{N:W_n} implies that there is a sequence of windows $W_{n}^B = \cup_{i=1}^{k_n} W_n^i$
given for each $n$ by a union of $m_n\times m_n$ sub squares $W_{n}^i$, $i=1,\cdots,k_n$,
such that $|W_{n}^B|/|W_n|\to 1$, $m_n=\hbox{O}(n^{\alpha})$ and the inter-distance
between any two neighboring sub squares is of order $n^{\eta}$ for some $4/(2 + \epsilon) < \eta < \alpha < 1$.
Let
\[
   \be_{n,\btheta^*}^B(\bbt) = \sum_{\bu\in X\cap W_n^B} \bphi_{n,\btheta^*}(\bu;\bbt) -
   \int_{W_n^B} \bphi_{n,\btheta^*}(\bu;\bbt) \lambda(\bu;\bbt)\dd\bu
   \equiv \sum_{i=1}^{k_n} \be_{n,\btheta^*}^{B,i}(\bbt),
\]
where
\[
   \be_{n,\btheta^*}^{B,i}(\bbt) = \sum_{\bu\in X\cap W_{n}^i}
   \bphi_{n,\theta^*}(\bu;\bbt) - \int_{W_{n}^i} \bphi_{n,\btheta^*}(\bu;\bbt) \lambda(\bu;\bbt)\dd \bu.
\]
Define
\[
   \tilde \be_{n,\btheta^*}^{B}(\bbt) = \sum_{i=1}^{k_n} \tilde \be_{n,\btheta^*}^{B,i}(\bbt),
\]
where the $\tilde \be_{n,\btheta^*}^{B,i}(\bbt)$'s are independent and for each
$i$ and $n$, $\tilde \be_{n,\btheta^*}^{B,i}(\bbt)$ is distributed as $\be_{n,\btheta^*}^{B,i}(\bbt)$.
Let $\bar \bsig^B_n = |W_n^B|^{-1}\Var \be_{n,\btheta^*}^B(\bbt^*)$ and
$\tilde \bsig_n^{B} = |W_n^B|^{-1}\Var \tilde \be_{n,\btheta^*}^{B}(\bbt^*)$.
We need to verify the following results:
\begin{enumerate}
\renewcommand{\theenumi}{\arabic{enumi}}
\renewcommand{\labelenumi}{S\theenumi}
\item \label{s:S1}
   $||\tilde \bsig_n^{B} - \bar \bsig^B_n||_M \to 0$ and $||\bar \bsig_n^{B} - \bar \bsig_n||_M \to 0$
   as $n\to\infty$, 
\item \label{s:S2}
   $|W_n^B|^{-1/2} \tilde \be_{n,\ta^*}^{B}(\bbt^*) \left( \tilde \bsig_n^{B} \right)^{-1/2}$
   is asymptotically standard normal,
\item \label{s:S3}
   $|W_n^B|^{-1/2} \be_{n,\btheta^*}^B(\bbt^*) \left( \tilde \bsig_n^{B} \right)^{-1/2} $
   has the same asymptotic distribution as\\
   $|W_n^B|^{-1/2} \tilde \be_{n,\btheta^*}^{B}(\bbt^*) \left( \bar \bsig_n^{B} \right)^{-1/2}$,
\item \label{s:S4}
   $\| |W_n^B|^{-1/2} \be_{n,\btheta^*}^B(\bbt^*) - |W_n|^{-1/2} \be_{n,\btheta^*}(\bbt^*)\|$
   converges to zero in probability.
\end{enumerate}
S\ref{s:S1} [C\ref{c:paircorrelation}, C\ref{c:phibounded}, N\ref{N:W_n}]:
  This follows from the proof of Theorem 2 in Guan and Loh (2007).\\[\bsl]
S\ref{s:S2} [C\ref{c:paircorrelation}, C\ref{c:phibounded}, N\ref{N:cumulant}]:
  Conditions C\ref{c:paircorrelation}, C\ref{c:phibounded} and N\ref{N:cumulant}
  imply $\EE[ \tilde \be_{n,\btheta^*}^{i}(\bbt)^4]$ is bounded \cite[see the proof of Lemma 1 in][]{guan:loh:07}.
  Thus, S2 follows from an application of Lyapunov's central limit theorem.\\[\bsl]
S\ref{s:S3} [N\ref{N:mixing}]:
  this follows by bounding the difference between the characteristic functions of
  $|W_n^B|^{-1/2} \be_{n,\btheta^*}^B(\bbt^*)$ and
  $|W_n^B|^{-1/2} \tilde \be_{n,\btheta^*}^{B}(\bbt^*)$ using techniques in~\cite{ibramigov:linnik:71}
  and secondly applying the mixing condition N\ref{N:mixing}, see also \cite{guan:sherman:calvin:04}.\\[\bsl]
S\ref{s:S4} [C\ref{c:rhobounded}-C\ref{c:phibounded}, C\ref{c:Inposdef}, N\ref{N:W_n}]:
  Recall that $|W_n^B|/|W_n|\to 1$ due to N\ref{N:W_n}. By C\ref{c:Inposdef} we only need
  to show $\Var\big[\be_{n,\theta^*}(\bbt^*) - \be_{n,\btheta^*}^B(\bbt^*) \big]/|W_n| \to 0$.
  This is implied by conditions C\ref{c:rhobounded}-C\ref{c:phibounded}
  and $|W_n^B|/|W_n|\to 1$.\\[\bsl]


\end{document}